\magnification = \magstep 1
\overfullrule=0pt
\font\twelvebf=cmbx12
\font\ninerm=cmr9
\nopagenumbers
\overfullrule=0pt
\line{\hfil CCNY-HEP 97/5}
\line{\hfil RU-97-3-B}
\line{\hfil SNUTP 97-054}
\line{\hfil March 1997}
\vskip .3in
\centerline{\twelvebf Planar Yang-Mills theory: Hamiltonian, regulators}
\centerline{\twelvebf and mass gap}
\vskip .3in
\baselineskip=14pt
\centerline{\ninerm DIMITRA KARABALI}
\vskip .05in
\centerline{Physics Department}
\centerline{Rockefeller University}
\centerline{New York, New York 10021}
\centerline{karabali@theory.rockefeller.edu}
\vskip .3in
\centerline{\ninerm CHANJU KIM}
\vskip .05in
\centerline{Center for Theoretical Physics}
\centerline{Seoul National University}
\centerline{151-742 Seoul, Korea}
\centerline{cjkim@ctp.snu.ac.kr}
\vskip .3in
\centerline{\ninerm V.P. NAIR}
\vskip .05in
\centerline{Physics Department}
\centerline{City College of the City University of New York}
\centerline{New York, New York 10031}
\centerline{vpn@ajanta.sci.ccny.cuny.edu}
\vskip .3in
\baselineskip=14pt
\centerline{\bf Abstract} We carry out the Hamiltonian analysis of non-Abelian gauge
theories in (2+1) dimensions in a gauge-invariant matrix parametrization of the fields.
A detailed discussion of regularization issues and the construction of the  renormalized
Laplace operator on the configuration space,  which is proportional to the kinetic
energy, are given.  The origin of the mass gap is analyzed and the lowest eigenstates of
the kinetic energy are explicitly  obtained;  these have zero charge and exhibit a mass
gap . The nature of the corrections due to the potential energy, the possibility of an
improved perturbation theory and a  Schrodinger-like equation for the states are also
discussed.
\vfill\eject

\footline={\hss\tenrm\folio\hss}
\def\bp {\bar p}
\def\ba {\bar{a}}
\def\bG {\bar{G}}
\def\bA {\bar{A}}
\def\bE {\bar{E}}
\def\bx {\bar{x}}
\def\by {\bar{y}}
\def\bu {\bar{u}}
\def\bv {\bar{v}}
\def\bw {\bar{w}}
\def\vk {\vec{k}}
\def\vx {{\vec x}}
\def\vz {\vec{z}}
\def\vy{\vec{y}}
\def\vv {\vec{v}}
\def\vu {\vec{u}}
\def\vw {\vec{w}}
\def\btheta {\bar{\theta}}
\def\dag {\dagger}
\def\del {\partial}
\def\bdel{\bar{\partial}}
\def\a {\alpha}
\def\b {\beta}
\def\e {\epsilon}
\def\d {\delta}
\def\s {\sigma}
\def\o {\omega}
\def\D {\Delta}
\def\bz {\bar{z}}
\def\12 {{\textstyle {1 \over 2}}}
\def\bG {\bar{G}}
\def\A {{\cal A}}
\def\C {{\cal C}}
\def\H {{\cal H}}
\def\O {{\cal O}}
\def\G {{\cal G}}
\def\F {{\cal F}}
\def\vf {\varphi}
\def\S {{\cal S}}
\def\ra {\rangle}
\def\la {\langle}
\def\Tr {{\rm Tr}}

\def\bV{{\bar V}}
\def\bD {{\bar D}}
\def\bA {{\bar A}}
\def\V {{\cal V}}
\baselineskip =18pt
\noindent{\bf 1. Introduction}
\vskip .1in Non-Abelian gauge theories are central to our current understanding of
physical phenomena. The perturbative analysis of such theories is fairly well understood
by now, having been extensively developed over the last three decades. Many of the
nonperturbative aspects are  also more or less understood at a qualitative level.
However, it is fair to say that, as of now, we do not have calculational techniques or
detailed understanding regarding nonperturbative phenomena in non-Abelian gauge
theories, eventhough there has recently been significant progress regarding the
nonperturbative aspects of supersymmetric gauge theories [1]. Recently we have analyzed
Yang-Mills theories in two spatial dimensions, in particular the question of how the
mass gap could be generated [2]. The motivation for considering the case of two spatial
dimensions is that it may capture some features of the more realistic case of three
dimensions (in this connection, see also refs.[3]), yet, at the same time, it would be
mathematically simpler to analyze since there are many known exact results about
two-dimensional field theories. Indeed, in our approach, in a Hamiltonian analysis we
are able to use a number of results from two-dimensional conformal field theory.  An
additional motivation is that there is at least one interesting physical situation,
viz., the high temperature phase of Chromodynamics and associated magnetic screening
effects, to which the (2+1)-dimensional theory can be directly applied.

Our approach was to carry out a Hamiltonian analysis utilising the geometrical
properties of the space of gauge-invariant field configurations $\C$. This configuration
space $\C$ is infinite-dimensional and the construction of a metric, volume element,
Laplacian, etc. requires appropriate regularization.  Regulators were used in arriving
at the various results presented in [2], although not all calculations were done within
a single regularization scheme.  In this paper, regularization issues are treated in
much greater detail; all calculations are done with essentially the same regulator
eliminating the possibilities of conflicts among different regulators used for different
calculations. The basic results are, of course, unchanged. We also carry out the
construction of the first excited eigenstate of the kinetic energy in detail.  (This was
only briefly indicated in [2].) The nature of the corrections due to the potential
energy term is also analyzed. Comparison of some results with the Abelian case and the
possibility of doing an ``improved" perturbation theory are additional new results in
this paper.

In the next section we give an outline of the main argument. The purpose of this is to
identify pieces of the calculations which need more careful regularized treatment. This
also serves to give a perspective on the fairly technical regularization issues
discussed in the subsequent sections. We introduce the matrix parametrization of fields
and obtain the volume element of $\C$ in terms of a Wess-Zumino-Witten (WZW) action for
a hermitian matrix. The wavefunctions can be taken as functions of the current of this
WZW theory and the arguments from conformal field theory which lead to this conclusion
are reviewed. The kinetic energy term of the Hamiltonian is proportional to the
Laplacian on the configuration space $\C$. We discuss the construction of this operator
and give the arguments for understanding how a mass gap arises.

In section 3 we define the basic regularization scheme. The volume element of $\C$,
adjointness properties of certain operators with the Haar measure for hermitian
matrices, self-adjointness of the kinetic energy and consistency with the Yang-Mills
(YM) equations are analyzed with this regulator. This is a fairly technical section with
detailed calculations, justifying many of the steps in our arguments. However, the later
sections can be read somewhat independently of this section.

An expression for the
 kinetic energy is obtained as an operator on functionals of currents in section 4.  In
section 5 we discuss some aspects of the Abelian theory; this is needed to clarify the
interpretation of some of the results in section 6.  The construction of the ground
state and the  first excited state of the kinetic energy operator, utilising the
regulated expressions of the previous sections is carried out in section 6. The question
of how self-energy subtractions can be done at any chosen scale and some issues related
to the choices of energy scales are also discussed. The effects of adding the potential
energy term are dicussed in section 7.

In section 8, we show that one can do an ``improved" perturbation theory where some of
the terms in the measure for integration over the configuration space are treated
exactly while other terms are expanded perturbatively. We show that this is consistent
with the self-adjointness of the Hamiltonian and also incorporates the mass gap. The
picture which emerges is as follows. One has ``constituent bosons" which carry
non-Abelian charge and behave as massive particles but which are interacting and get
bound into states of zero charge. The expansion scheme of section 8 can potentially be
used for a systematic analysis of higher excited states. This is currently under
investigation.

We conclude with a discussion comparing our results to the electric field representation
as well as estimating the significance of the potential energy term.

\vskip .1in
\noindent{\bf 2. Outline of the main argument}
\vskip .1in In this section we give an outline of the main argument before zeroing in on
specific pieces of the calculations which need more elaborate analysis using regulators. 

We shall discuss an $SU(N)$-gauge theory.  As is convenient for a Hamiltonian
formulation, we shall work in the $A_0 =0$ gauge. The gauge potentials can be written as
$A_i = -i t^a A_i ^a$, $i=1,2$, where $t^a$ are hermitian 
$(N \times N)$-matrices which form a basis of the Lie algebra of $SU(N)$ with
$[t^a, t^b ] = i f^{abc} t^c,~~{\rm {Tr}} (t^at^b) = {1 \over 2} \delta ^{ab}$. 
${\A}$ will denote the set of all gauge potentials $A_i ^a$. Gauge transformations act
on $A_i$ in the standard way, $A_i \rightarrow A_i ^g$, where
$$ A_i^{(g)} = g A_i g^{-1} ~-  \partial_i g g^{-1} \eqno(2.1)
$$ and $g( \vec x ) \in SU(N)$ The gauge group ${\G}_*$ is defined by
$$ {\G}_* = \left\{ {\rm set ~of ~all} ~g(\vec x ): {\bf R}^2 \rightarrow SU(N),
~~g\rightarrow 1 ~{\rm as} ~\vert \vec x \vert \rightarrow \infty \right\}
\eqno(2.2)
$$  The space of gauge-invariant field configurations is 
$$
\C ={\A}/{\G}_* \eqno(2.3)
$$ 

The basic strategy adopted in papers [2] was to formulate the calculation, as much as
possible, in terms of the geometry of $\C$. (For discussions on the geometry of $\C$,
see ref. [4].) Specifically we shall need a metric and volume element on $\C$ and
eventually also the Laplace operator
$\Delta$ on $\C$. The YM Lagrangian in the $A_0 =0$ gauge is given by
$$
 L=  \int d^2x~ \left[ { e^2 \over 2 } {{\partial A_i^a} \over {\partial t}}  {{\partial
A_i^a}
\over {\partial t}}  -  {1 \over {2 e^2}} B^a B^a \right] \eqno(2.4)
$$  where $B^a= \12 \epsilon_{ij} (\partial_i A_j^a - \partial_j A_i^a +f^{abc}A_i^b
A_j^c)$. By comparison of the kinetic term with the standard point-particle Lagrangian
$L = \12 g_{\mu \nu} {dq^{\mu} \over dt} {dq^{\nu}
\over dt}$, we see that the metric on the space of potentials $\A$ which is appropriate
for YM theory is 
$$  ds^2_{\A}~ =~ \int d^2x~ \delta A^a_i \delta A^a_i  \eqno(2.5)
$$  This is the starting point for calculations on $\C$. A good parametrization of the
fields $A_i^a$ which allows for explicit calculations is a first step in the reduction
of this metric to the gauge-invariant configuration space $\C$.  (There are many
different parametrizations which have been studied; for some other parametrizations, see
ref. [5].) We shall combine the spatial coordinates
 $x_1 ,x_2$ into the complex combinations
$z=x_1 -ix_2,~{\bar z} =x_1+ix_2$; correspondingly we have
$A\equiv A_{z} = {1 \over 2} (A_1 +i A_2), ~~  {\bar A}\equiv A_{\bar{z}} = {1
\over 2} (A_1 -i A_2) = - (A_z)^{\dagger}$. The parametrization we use is given by
$$  A_z = -\partial_{z} M M^{-1},~~~~~~~~~~~~~~~~ A_{\bar{z}} = M^{\dagger -1}
\partial_ {\bar{z}} M^{\dagger}
\eqno(2.6) 
$$ Here $M,~M^\dagger$ are complex $SL(N,{\bf {C}})$-matrices  (for an
$SU(N)$-gauge theory; for group $G$, $M,~M^{\dagger}$ belong to $G^{\bf C}$, the
complexification of $G$). Such a parametrization  is standard  in many discussions of
two-dimensional gauge fields. We can define Green's functions
$G,~\bar{G}$ for $\partial,~\bar{\partial}$ by
$$\eqalign{ &\partial _x G(\vec x, \vy) ~= { \bar \partial}_x \bar{G} (\vx,\vy) = \delta
^{(2)} (\vx-\vy) \cr &G(\vx, \vx') = {1 \over {\pi (\bar{z} - \bar{z}')}}
~~~~~~~~~\bar{G} (\vx, \vx') = {1 \over {\pi (z-z')}} \cr}
\eqno(2.7)
$$ We have chosen the boundary condition $G,~\bar{G} \rightarrow 0$ as $\vert
\vx-\vx' \vert \rightarrow \infty$. For any gauge potentials $A,~\bA$,  one can easily
check that a choice of $M,~M^\dagger$ is given by
$$\eqalign{ M(\vx) &= 1 - \int G(\vx, \vz_1 ) A(\vz_1) +\int G(\vx,\vz_1)A(\vz_1)
G(\vz_1,\vec{z}_2)
 A(\vec{z}_2) - ...\cr &=~ 1 -\int_y  D^{-1}(\vx,\vy) A (\vy)\cr M^\dag (\vx) &= ~1 -
\int_y
\bA  (\vy) \bD ^{-1} (\vy,\vx)\cr }\eqno(2.8)
$$ Here $D=\partial +A,~\bD =\bdel +\bA$ are covariant derivatives. (There may be many
choices for $M,~M^\dagger$; we shall discuss this a little later.) From the definition
(2.6), it is clear that the gauge transformation (2.1)  is expressed in terms of $M,
~M^\dagger$ by
$$ M\rightarrow M^{(g)}=gM,~~~~~~~~~~~M^{\dagger (g)}=M^\dagger g^{-1} \eqno(2.9)
$$ for $g(\vx) \in SU(N)$. In particular, if we split $M$ into a unitary part $U$  and a
hermitian part $\rho$ as $M=U\rho$, then $U$ is the `gauge part',  so to speak; it can
be removed by a gauge transformation and $\rho$  represents the gauge-invariant degrees
of freedom. Alternatively,  we can use $H=M^\dagger M =\rho^2$ as the gauge-invariant
field  parametrizing $\C$. Since $M\in SL(N,{\bf {C}}),~\rho$, and hence $H$,  belong to
$SL(N,{\bf {C}})/SU(N)$.

In terms of the parametrization (2.6), the metric (2.5) can be written as
$$\eqalign{ ds^2_{\A}~ &=~ \int d^2x~ \delta A^a_i \delta A^a_i ~=  -8 \int \Tr (\delta
A_z \delta A_{\bz} )\cr &= 8 \int \Tr \left[ D(\delta M M^{-1}) \bD (M^{\dag -1}\delta
M^\dag )
\right] \cr}\eqno(2.10)
$$ where the covariant derivatives $D,\bD$ are in  the adjoint representation. Two
remarks about this metric are in order. This is a standard Euclidean metric in terms of
$A$'s and hence the corresponding volume element $d\mu (\A)$ for
$\A$ is the standard Euclidean one, i.e., $d\mu (\A)=[dAd\bar{A}]= \prod_{x,a} dA^a(\vx)
dA^a (\vx)$. Secondly, this is a K\"ahler metric. Evidently
$$\eqalign{ & ds^2_{\A} = \delta _{A} \delta _{\bar{A}} W \cr & W=-8 \int Tr (A
\bar{A}) + f(A) + \bar{f} (\bar{A}) \cr}
\eqno(2.11)
$$ The K\"ahler potential $W$ is defined, as usual, only upto the addition of a purely
$A$-dependent function $f(A)$ and a purely $\bar{A}$-dependent function
$\bar{f}(\bar{A})$. (It is possible to choose $f,~\bar{f}$ such that $W$ is
gauge-invariant which is nice but not particularly relevant to our discussion.)

The matrices $M,~M^\dag$ are elements of $SL(N,{\bf {C}})$ and  we have the
Cartan-Killing metric for $SL(N,{\bf {C}})$, viz., $ds^2~=8 {\Tr}(\delta M
M^{-1}~M^{\dagger -1} \delta M^\dagger )$. For $SL(N,{\bf {C}})$-valued fields, we thus
have
$$ ds^2_{SL(N,\bf {C})}  = 8 \int \Tr [(\delta M M^{-1}) (M^{\dagger -1} \delta
M^{\dagger})] 
\eqno(2.12)
$$ We denote the corresponding volume element, the Haar measure, by 
$d\mu (M, M^\dagger )$. From Eq.(2.10) we can see that $d\mu (\A ) =\det (D \bD ) d\mu
(M,M^\dagger )$.

The volume element for $SL(N,{\bf C})$ is of the form
$$\eqalign{ dV(M,M^{\dagger}) \propto & \epsilon _{a_1...a_n}  (dM M^{-1})_{a_1}
\wedge ...  \wedge (dMM^{-1})_{a_n}  \cr
 \times & \epsilon _{b_1...b_n} (M^{\dagger -1} d M^{\dagger})_{b_1} \wedge ...
\wedge  (M^{\dagger -1} d M^{\dagger})_{b_n}  \cr}
\eqno(2.13)
$$  where $n={\rm dim} G= N^2 -1$. (We use proportionality relationship, there are some
constant numerical factors which are irrelevant for our discussion.)
 By direct substitution of $M = U\rho$, Eq.(2.13) becomes
$$\eqalign{ dV(M, M^{\dagger}) \propto  & \epsilon _{a_1...a_n} (d\rho \rho^{-1} +
\rho^{-1} d\rho)_{a_1} \wedge ... \wedge (d\rho \rho^{-1} + \rho^{-1} d\rho)_{a_n}  \cr
\times & \epsilon _{b_1...b_n} (U^{ -1} d U)_{b_1} \wedge ... \wedge (U^{ -1} d
U)_{b_n}  \cr
\propto & \epsilon _{a_1...a_n} (H^{-1}dH)_{a_1} \wedge ...  \wedge (H^{-1}dH)_{a_n}  
d\mu (U) \cr}
\eqno(2.14)
$$ Here $d\mu (U)$ is the Haar measure for $SU(N)$. Notice that
$$
 \epsilon _{a_1...a_n} (H^{-1}dH)_{a_1} ... (H^{-1}dH)_{a_n}   = \det r \prod d\vf ^a
\eqno (2.15)
$$ where $H^{-1}dH = d\vf ^a r_{ak} (\vf) t_k$. We parametrize $H$ in terms of the real
parameters $\vf ^a$. Upon taking the product over all points, $d\mu (U)$ gives the
volume of $\G _*$ and thus
$$\eqalign{ d\mu (M, M^{\dagger}) = & = \prod_{x} dV(M, M^{\dag}) = \prod \det r [d \vf]
~ vol (\G _*) \cr = & d\mu (H)  ~vol (\G _*) \cr}
\eqno(2.16)
$$
$d\mu (H)  = \prod _{x}  \det r [d\vf ^a]$ is the Haar measure for hermitian
matrix-valued fields and is the volume element associated with
$$  ds^2 ~= 2 \int ~\Tr(H^{-1}~\delta H )^2  \eqno(2.17)
$$  (This metric can also be derived directly by ``reduction" of Eq.(2.12) to the coset
space $SL(N,{\bf C}) /SU(N)$.) From Eq.(2.16) we have
$$ {{d\mu(M, M^{\dagger})} \over vol ({\G}_*)} = d\mu (H) \eqno(2.18)
$$  The volume element for $\C $ is now obtained as
$$
\eqalign{ d\mu ({\C}) &= {d\mu ({\A})\over vol({\G}_*)}
 = {[dA_z dA_{\bar{z}}]\over vol({\G}_*)} \cr &= (\det D_z D_{\bar{z}}) {d\mu  (M,
M^{\dagger})\over vol({\G}_*)} ~= (\det D \bD ) d\mu (H)\cr}\eqno(2.19) 
$$  The problem is thus reduced to the calculation of the determinant of the
two-dimensional operator $D\bD$. This is well known [6]. We get
$$  (\det D \bD) ~= \left[ {{\det ' \del \bdel } \over \int d^2 x} \right] ^{{\rm dim}
G} ~ \exp \left[ 2c_A ~\S (H) \right]\eqno(2.20)
$$  where $c_A \delta^{ab} = f^{amn}f^{bmn}$ and $\S (H)$ is the  Wess-Zumino-Witten
(WZW) action for the hermitian matrix field $H$ given by [7]
$$  {\S} (H) = {1 \over {2 \pi}} \int \Tr (\partial H \bar{\partial} H^{-1}) +{i
\over {12 \pi}} \int \epsilon ^{\mu \nu \alpha} \Tr ( H^{-1} \partial _{\mu} H H^{-1}
\partial _{\nu}H H^{-1} \partial _{\alpha}H) \eqno(2.21) 
$$ The calculation of the $(\det D \bar{D})$ is most easily done as follows. Defining
$\Gamma = \log~\det D\bD$, we find
$$ {\delta \Gamma \over \delta \bA^a}~= -i~\Tr\left[ \bD^{-1}(\vx,\vy) 
T^a\right]_{\vy\rightarrow \vx} \eqno(2.22)
$$
$(T^a)_{mn}=-if^a_{mn}$ are the generators  of the Lie algebra in the adjoint
representation. The coincident-point limit of $\bD^{-1}(\vx,\vy)$ is, of course,
singular and needs regularization. Since $ d\mu (\C )$ must be gauge-invariant, a
gauge-invariant regularization is appropriate here. With a gauge-invariant regulator, as
we shall see in the next section,
$$
\Tr \left[ \bD^{-1}_{reg}(\vx,\vy) T^a \right]_{\vy\rightarrow \vx}~= {2c_A \over \pi}
\Tr \left[ (A -M^{\dag -1} \partial M^\dag )t^a\right] \eqno(2.23)
$$ Using this result in Eq.(2.22), and with a similar result for the variation of 
$\Gamma$ with respect to $A^a$, and integrating we get $\Gamma = 2c_A \S (H)$.

The calculation in Eq.(2.23) is essentially the anomaly calculation in two dimensions
and the result is quite robust; different regulators, such as covariant point-splitting,
Pauli-Villars, etc., lead to the same result so long as gauge invariance is preserved. 

We now have the result, upto an irrelevant constant factor [8,9],
$$ d\mu (\C )~= d\mu (H)  ~e^{2c_A \S (H)} \eqno(2.24)
$$  The inner product for physical states is given by
$$
\la 1 \vert 2\ra = \int  d\mu (H)  e^{2c_A \S (H)}~\Psi_1^* (H) \Psi_2 (H)
\eqno(2.25)
$$  This formula shows that all matrix elements in $(2+1)$-dimensional
$SU(N)$-gauge theory  can be evaluated as  correlators of the $SL(N,{\bf {C}})/SU(N)$-
WZW model. (For a general gauge group $G$, we will have a
$G^{\bf {C}}/G$-WZW model, $G^{\bf {C}}$ being the complexification of $G$.)

There is an interesting point of comparison between Eq.(2.24) and  two-dimensional
Euclidean YM theory. First of all, notice that the  ``total volume" of $\C$ as given by
$\int d\mu (\C )$ is the partition function for a Euclidean two-dimensional hermitian
WZW model. This can be explicitly evaluated as [8,9]
$$
\int d\mu (H) ~\exp \left[ 2 c_A \S (H) \right] ~= 
\left[ {{\det ' \del \bdel } \over \int d^2 x} \right] ^{-{\rm dim} G} \eqno(2.26)
$$ The ``total volume" of $\C$, so defined, is finite. Indeed, if we retain the
determinantal factors from Eq.(2.20), we find that $\int d\mu (\C ) =1$. (We should keep
in mind that there is still a regularization implicit in this statement.)

Since we are dealing only with $A_z, A_{\bar z}$, the gauge-invariant measure
$d\mu (\C )$ is identical to what is needed for the functional integral of
two-dimensional Euclidean YM theory. The partition function for that theory has been
evaluated on arbitrary Riemann surfaces by different methods and is given by [10]
$$\eqalign{ Z(g)&= \int d\mu (\C )~ \exp \left[ -{1\over 4 g^2}\int d^2x~F^a_{\mu\nu}
F^a_{\mu\nu}\right]\cr &=\sum_{R} d_R^{2-2G}~\exp\left(- \12 g^2  c_R ~A\right)\cr}
\eqno(2.27)
$$ Here $g$ is the two-dimensional coupling constant, $G$ is the genus of the Riemann
surface and $A$ is its area. The summation is over all the irreducible representations
of $SU(N)$; $d_R$ is the dimension and $c_R$ is the quadratic Casimir of the
representation $R$. Given this result, we may define $\int d\mu (\C )$ as the limit of
$Z(g)$ for $g\rightarrow \infty$. We see that only the identity representation survives
on the right hand side of Eq.(2.27) giving $\int d\mu (\C )=
\lim_{g\rightarrow \infty} Z(g) =1$, which is consistent with what we found.

Some properties of the hermitian WZW-model will be relevant to our discussion. These can
be obtained by comparison  with the $SU(N)$-model defined by $e^{k
\S (U)},~ U(\vx)\in SU(N)$. The quantity which corresponds to $e^{k\S (U)}$ for the
hermitian model is $e^{(k+2c_A)
\S (H)}$. The hermitian analogue of the renormalized level $\kappa = (k+c_A)$ of the
$SU(N)$-model is $-(k+c_A)$. Correlators can be calculated from the
Knizhnik-Zamolodchikov equation. Since the latter involves only the  renormalized level
$\kappa$, we see that the correlators of the  hermitian model (of level $(k+2c_A)$ ) can
be obtained from the correlators of  the
$SU(N)$-model (of level $k$ ) by the analytic continuation 
$\kappa \rightarrow -\kappa$. For the $SU(N)_k$-model there are the  so-called
integrable representations whose highest weights are limited  by $k$ (spin $\leq k/2$
for $SU(2)$, for example). Correlators involving  the nonintegrable representations
vanish. For the hermitian model the  corresponding statement is that the correlators
involving nonintegrable  representations are infinite [8,9].

In our case, $k=0$, and we have only one integrable representation corresponding to the
identity operator (and its current algebra descendents). The matrix elements of the
gauge theory being correlators of the hermitian WZW-model, we have the result that all
wavefunctions  of finite inner product and norm are functions of  the current
$$  J_a(\vx) = {c_A \over \pi } \left( \partial H~H^{-1} \right)_a (\vx) ~= {c_A
\over \pi} \left[ iM^{\dag} _{ ab} (\vx) A_b(\vx) +  (\partial M^\dag ~M^{\dag -1})_a
(\vx)
\right] \eqno(2.28)
$$  where $M^{\dag} _{ab}= 2 \Tr (t^a M^\dag t^b M^{\dag -1} )$ is the adjoint
representation of $M^\dag$. (This restriction from conformal field theory can be evaded
if wavefunctions are so chosen as to have a growing exponential which compensates for
the factor $e^{2 c_A \S (H)}$ in (2.25). However such wavefunctions will have infinite
expectation values for the potential energy term $\int B^2$ and can therefore be ruled
out.)

It is instructive to consider how this infinite value arises for correlators of
nonintegrable representations. Consider the four-point function for four $H$'s taken in
the fundamental representation, i.e., $(N \times N)$-matrices, for the hermitian model
of level number $(k+2c_A)$. This is given by [9,11]
$$\eqalignno{
  <H_{i_1 j_1}(1) H^{-1} _{i_2 j_2} (2) H_{i_3 j_3} (3) H^{-1} _{i_4 j_4} (4) > = & \sum
_{p=0,1} \lambda _p (\bar{M} ^p _1 \delta _{i_1 i_2} \delta _{i_3 i_4} + \bar{M} ^p _2
\delta _{i_1 i_4} \delta_{i_2 i_3} )  &{} \cr 
  & \times (M ^p _1 \delta _{j_1 j_2} \delta _{j_3 j_4} +M ^p _2 \delta _{j_1 j_4}
\delta _{j_2 j_3} )   &(2.29) \cr
  M^p_A = (z_{13} z_{24}) ^{{{N^2-1} \over {N(N+k)}}} {\F}^{(p)} _A (x), ~~~~~~~~~~~~~ &
x={{z_{12}z_{34}} \over {z_{13} z_{24}}},~~~~~~~~~~z_{ij} \equiv z_i -z_j   &{} \cr}
$$
$\lambda _0 =1$ and $\lambda _1 = h(-N-k)$ where
$$ h(\kappa)= {1 \over N^2} {{\Gamma({N-1 \over \kappa}) \Gamma({N+1 \over
\kappa})\Gamma ^2({k \over \kappa})} \over {\Gamma({k+1 \over
\kappa})\Gamma({k-1 \over \kappa})\Gamma ^2 ({N \over \kappa})}} \eqno(2.30)
$$  (We have used the fact that $c_A = N$ for $SU(N)$.)As $k \rightarrow 0$, we find
$\lambda _1 = {1 \over {N^2 -1}}$. The chiral blocks ${\cal{F}} ^{(0) }_A$ are
nonsingular as $k \rightarrow 0$. The
${\cal{F}} ^{(1)} _A$ are given by
$$\eqalign{
\F^ {(1)} _1 & = [x(1-x)]^{{-1 \over {N(k+N)}}} F \bigl( -{(N-1) \over {k+N}}, - {(N+1)
\over {k+N}}, {k \over {k+N}}, x \bigr) \cr
\F^ {(1)} _2 & = -N [x(1-x)]^{{-1 \over {N(k+N)}}} F \bigl( -{(N-1) \over {k+N}}, -
{(N+1) \over {k+N}}, -{N \over {k+N}}, x \bigr) \cr}
\eqno(2.31)
$$ where $F(\alpha, \beta, \gamma,x)$ is the hypergeometric function. The hypergeometric
function has simple poles at $\gamma =0, -1, -2,...$ (which are the same as for the
Eulerian gamma function [12]) and so, as $k \rightarrow 0$, these chiral blocks become
infinite for any value of $x$. Notice that this is not a spacetime singularity, or a
regularization problem, it is a singularity in the coupling constant $k$. This is in
agreement with our statements since the fundamental representation is nonintegrable for
$k \rightarrow 0$.

Wavefunctions, as we have argued, are functions of the current (2.28). From the physical
point of view of accounting for all the gauge-invariant degrees of freedom this is not a
limitation since the Wilson loop operator can be written in terms of the current as
$$ W(C)~=~ \Tr ~P ~e^{-\oint_C (Adz+\bA d{\bar z})}~=
 \Tr ~P~e^{(\pi /c_A)\oint_C J }\eqno(2.32)
$$ and, at least in principle, all gauge-invariant functions of $(A, \bA )$ can be
constructed  from $W(C)$.

We now turn to the construction of the kinetic energy term which is proportional to the
Laplacian on $\C$. First consider the change of variables from $A, \bA$ to $M,
M^{\dag}$. Parametrizing $M, M^{\dag}$ in terms of $\theta ^a (\vx),~\btheta ^a (\vx)$
respectively, we can write
$$ M^{-1}\delta M~ = \delta \theta^a R_{ab}(\theta ) t_b, ~~~~~ 
\delta M^\dag M^{\dag -1} ~= \delta {\bar \theta}^a R^*_{ab}({\bar \theta})t_b
\eqno(2.33)
$$ (These define $R_{ab},~R_{ab}^*$.) One can now write the electric field operators as
$$\eqalign{ E_k (\vx) = -{i \over 2} {\delta \over {\delta \bA _k (\vx)}} = {i \over 2}
M^{\dag} _{ak} (\vx) \int_y \bG (\vx,\vy) \bp _a (\vy) \cr
\bE_k (\vx) = -{i \over 2} {\delta \over \delta A _k (\vx)} =- {i \over 2} M _{ka} (\vx)
\int_y G (\vx,\vy) p _a (\vy) \cr} \eqno(2.34)
$$  where $M_{ab}= 2\Tr (t^a M t^b M^{-1})$ is the adjoint representation of $M$.
$p_a$ is the right-translation operator on $M$ and $\bp _a$ is the left-translation
operator on $M^{\dag}$, i.e.,
$$\eqalign{ [p_a(\vx), M(\vy)] ~&=M(\vy) (-it_a)~\delta (\vy-\vx)\cr [\bp_a (\vx), M^\dag
(\vy) ] ~&= (-it_a) M^\dag (\vy) ~\delta (\vy-\vx)\cr}\eqno(2.35)
$$ Explicitly
$$ p_a (\vx) = -i R^{-1} _{ab}(\vx) {\delta \over {\delta \theta ^b (\vx)}},~~~~~~~
\bp_a (\vx) = -i R^{* -1} _{ab}(\vx) {\delta \over {\delta \btheta ^b (\vx)}}
\eqno(2.36)
$$  The kinetic energy operator is given by
$$
 T= -{e^2 \over 2} \int_x {\d ^2 \over {\d A_k (\vx) \d \bA _k (\vx)}} = {e^2
\over 2} \int_x K_{ab} (\vx) (\bG \bp _a) (\vx) (G p _b) (\vx) \eqno(2.37)
$$  where $K_{ab} = M^{\dag}_{ak} M_{kb} = 2 \Tr (t^a H t^b H^{-1})$ and
$Gp_b(\vx)\equiv \int_y G(\vx, \vy)p_b(\vy)$, etc. This expression is still defined on
$\A$. With a splitting $M=U \rho$ we can write
$$
\eqalign{ & p_a = (1+\rho)^{-1} _{ab} (\a _b + I_b) \cr & \bp _a = [\rho ^2 (1 +
\rho)^{-1}]_{ab} \a _b - [\rho (1+\rho)^{-1}]_{ab} I_b \cr}
\eqno(2.38)
$$  where $\a _a$ generates right translations on $\rho$ and $I_a$ generates right
translations on $U$, i.e., $[\a _a (\vx), \rho (\vy)] = \rho (\vy) (-i t^a) \d
(\vy-\vx),~~ [I_a (\vx), U(\vy)]= U(\vy) (-it^a) \d (\vy-\vx)$ and $\rho _{ab} =2 \Tr
(t^a
\rho t^b \rho ^{-1})$. On functions which are gauge-invariant and hence independent of
$U$, the action of $I_a$ gives zero. The operator $T$, for functions on $\C$, is thus
given by Eq.(2.37) with $p_a,~\bp _a$ as in  Eq.(2.38), but with $I_a$ set to zero. When
$I_a$ is set to zero, one can easily check that
$$\eqalign{ [p_a(\vx),~H(\vy)]&= H(\vy) (-it^a)\d (\vy-\vx) \cr 
 p_a(\vx) &= -i r^{-1} _{ab} {\d \over {\d \vf ^b}} \cr 
 \bp _a(\vx) &= K_{ab}(\vx)p_b(\vx) \cr}
\eqno(2.39)
$$  ($\bp _a(\vx)$ generates left-translations on $H$.) This gives the construction of
$T$ as an operator on functions on $\C$. The expression (2.37) however needs
regularization. It is not manifestly self-adjoint. In checking self-adjointness one
encounters singular commutators such as $[\bG \bp _a(\vx),~ K_{ab}(\vx)]$. Further,
$p_a,~\bp _a$ are formally adjoints of each other with respect to the Haar measure $d\mu
(H)$. One needs to check whether this holds in an appropriately regularized version.

$T$ is proportional to the Laplacian on $\C$. One may, starting from the metric on $\A$,
directly construct it as well. We have
$$\eqalignno{  ds^2_{\A}~& = -8 \int \Tr (\delta A \delta \bA ) = \int g_{a{\bar
b}}(\vx,\vy) 
\delta\theta^a(\vx) \delta {\bar\theta}^b(\vy)~+~ h.c.&(2.40a)\cr
 g_{a{\bar b}}(\vx,\vy) ~&= 2 \int_{u,v} \partial_u [\delta (\vu-\vx)R_{ar}(\vu)]
M_{kr}(\vu) M^\dag_{sk}(\vv) {\bar \partial}_v [\delta (\vv-\vy)
R^*_{bs}(\vv)]&(2.40b)\cr}
$$  The Laplacian $\D$ on a complex manifold has the general form
$$
\D~= g^{-1}\left( \partial_{\bar a} g^{{\bar a}a}g \partial_a + 
\partial_a g^{a{\bar a}}g \partial_{\bar a}\right) \eqno(2.41)
$$  where $g=\det(g_{a{\bar a}})$. Using the metric components (2.40b) we find
$$\eqalign{  T~= -{e^2\over 2} \D = {e^2\over 4} \int_x e^{-2c_A \S (H)} \bigl[ &{\bar
G}\bp_a(\vx) K_{ab}(\vx) e^{2c_A\S (H)} Gp_b(\vx) ~+\cr & Gp_a(\vx) K_{ba}(\vx)
e^{2c_A\S (H)} {\bar G}\bp_b(\vx)\bigr]\cr}\eqno(2.42)
$$  Once again we obtain $T$ on $\C$ by setting $I$'s to zero in Eq.(2.38) for
$p_a,~\bp _a$. This expression has manifest self-adjointness since 
$$\eqalign{
\la 1|T |2\ra &= {e^2 \over 4} \int d\mu (H) e^{2 c_A \S (H)} \left[ {\overline {G p_a
\psi _1}} K_{ab} (G p_b \psi _2) + {\overline { \bG \bp _a \psi _1}} K_{ba}  (\bG
\bp _b \psi _2) \right]\cr & = \la T1|2\ra \cr}  \eqno(2.43)
$$  provided $p,~\bp$ are adjoints of each other with respect to $d\mu (H)$. Notice that
if we attempt to move $\bG \bp _a$ through $K_{ab} e^{2 c_A \S}$ to the right end we
encounter the singular commutator $[\bG \bp _a(\vx), K_{ab}(\vx)]$. Again expression
(2.42) needs to be regularized to show that it is the same as expression (2.37), thereby
proving self-adjointness of the latter form of
$T$.

The identity of the two expressions (2.37, 2.42) is obtained if, in terms of metric
components, we have $\del _{\ba} (g^{\ba a} g) =0$, so that all derivatives in Eq.(2.41)
can be moved to the right.  As we have noted in Eq.(2.11), the metric (2.40a) is a
K\"ahler metric. For a finite-dimensional K\"ahler metric, $\del _{\ba} (g^{\ba a} g)
=0$, as  can be checked directly. Showing the identity of expressions (2.37, 2.42) is
thus equivalent to proving $\del _{\ba} (g^{\ba a} g) =0$ for the infinite-dimensional
case we have. (This interpretation in terms of a K\"ahler property is clearly on $\A$,
i.e., before we set
$I_a$'s to zero in $p_a, ~\bp_a$.  There is no reason why $\C$ should be a K\"ahler
manifold.)

The regularization questions we have isolated so far have to do with the adjointness of
$p_a,~\bp _a$ with respect to $d\mu (H)$ and the equality of expressions (2.37, 2.42).
For the sake of completeness, we shall also recheck Eq.(2.23) in the calculation of the
determinant in the next section eventhough this is essentially the anomaly calculation.

As we have discussed before, it suffices to consider wavefunctions which are functions
of the current $J_a$. Therefore, before leaving this section, we shall evaluate the
action of $T$ on $J_a(\vx)$, which is the simplest case to consider. Using the
expression (2.28) for the current in terms of $M^{\dag}$ and $A$, we find
$$\eqalignno{
 T~J_a(\vx) &= -{e^2\over 2}\int d^2y {\delta^2 J_a(\vx)\over 
\delta \bA^b(\vy) \delta A^b(\vy)} ~={e^2c_A\over 2\pi} M^{\dag} _{am} \Tr \left[ T^m
\bD ^{-1}(\vy,\vx) 
\right]_{\vy \rightarrow \vx}&(2.44a)\cr  &= m ~J_a(\vx)&(2.44b)\cr}
$$  We encounter the same expression with the coincident-point limit as in the
calculation of $(\det D\bD)$. Using the same result as in that calculation, viz.,
Eq.(2.23), we get the result (2.44b). The validity of this particular  result is thus on
the same footing as the calculation of the volume element 
$d\mu (\C)$.

The Hamiltonian ${\cal{H}} = T + V$ can also be expressed entirely in terms of $H$ or
the current $J_a$ since the potential energy $V$ can be written as 
$$ V= \int {{B^a B^a} \over {2 e^2}} = {\pi \over {m c_A}} \int \bdel J_a \bdel J_a
\eqno(2.45)
$$
\vskip .1 in
\noindent{\bf 3. Regularization}
\vskip .1in We now consider the choice of a regulator. Any regulator we choose must, of
course, be gauge-invariant. It should also respect the ``holomorphic invariance". This
arises as follows. We have discussed the construction of $M, M^{\dag}$ for given
potentials $A, \bA$. The boundary condition $G,~\bG
\rightarrow 0$ as $|\vx-\vx'| \rightarrow \infty$ leads to the choice (2.7) for the
Green's functions. Even so, there are many possible choices for $M,~M^{\dag}$.  Eq.(2.6)
may be inverted as
$$\eqalign{ M(\vx) = & \bV (\bx) + \int A(\vz_1) M(\vz_1) G (\vz_1,\vx) \cr = & \bV (\bx)
+ \int A(\vz_1) \bV (\bz _1) G(\vz_1,\vx) +...  = (1+ \int A(\vz_1) \tilde{G} (\vz_1,\vx)
+..) \bV (\bx) \cr}
\eqno(3.1)
$$ where $\tilde{G} (\vz,\vx) = \bV (\bz ) G(\vz,\vx)) \bV ^{-1} (\bx)$. Thus a different
choice of the starting point of the iteration, viz., $\bV (\bx)$ rather than 1,
corresponds to $M \rightarrow M \bV (\bx)$ and $G(\vx, \vy) \rightarrow \bV (\bx) G(\vx,
\vy) \bV ^{-1} (\by)$. Since we have the same $A,~\bA$, clearly physical results must be
insensitive to this redundancy in the parametrization in terms of $M, M^{\dag}$; we must
require invariance under $M \rightarrow M \bV (\bx),~ M^{\dag} \rightarrow V(x)
M^{\dag},~ H \rightarrow V H \bV $ (and $G(\vx, \vy)
\rightarrow \bV (\bx ) G(\vx, \vy) \bV ^{-1} (\by),~ \bG (\vx,\vy) \rightarrow V(x) \bG
(\vx,\vy) V^{-1} (y)$ appropriately). We shall refer to this requirement as holomorphic
invariance. (To clarify the notation in the above discussion,
$V,~\bV$ depend only on the holomorphic coordinate
$x$ and the antiholomorphic coordinate $\bx$ respectively while the fields $M,~M^\dagger
,~A,~\bA $ in general depend on both $x,\bx$, indicated by the vector notation for the
arguments.)

Of course there are no antiholomorphic functions $\bV (\bx)$, except for $\bV$ being a
constant matrix, if we do not allow singularities in $\bV$ anywhere including spatial
infinity. Generally what happens in such a case is that $H$ can have singularities. The
location of these singularities can be changed by the transformation $H \rightarrow
VH\bV$. One can eliminate such singularities by defining $M,~M^{\dag}$ and hence $H$
separately on coordinate patches and using (anti) holomorphic matrices as transition
functions,  i.e., $M_1= M_2 {\bar V}_{12},$ etc., or $H_1 = V_{12} H_2 {\bar V}_{12}$ 
in terms of $H=M^\dag M$. Since this is an ambiguity of choice of field  variables, the
wavefunctions must be invariant under this. (The ambiguity in  the choice of $M$ or $H$
and the need for (anti)holomorphic transition  functions are related to the geometry of
$\A$ as a ${\G}_*$-bundle  over $\C$ and the Gribov problem [13]. For a discussion of
these issues,  see reference [2].)

We can easily check the holomorphic invariance of  the various expressions we have. From
the Polyakov-Wiegman formula [6] 
$$ \S (h_1 h_2) = \S (h_1) + \S (h_2) - {i \over \pi} \int \Tr ( \bdel h_1 h^{-1}_1
h_2^{-1} \del h_2) \eqno(3.2)
$$ we can see that $\S (V~H~\bV) =\S (H)$; this is the Kac-Moody symmetry of the
WZW-model. $d\mu(H)$ is easily checked to be invariant. When $M
\rightarrow M~\bV,~p_a \rightarrow \bV _{ab} p_b$ and, similarly, $\bp _a
\rightarrow V_{ab}~\bp _b$ for $M^{\dag} \rightarrow V~M^{\dag}$, as may be checked from
Eq.(2.35). With the change $G \rightarrow \bV~G~\bV ^{-1},~\bG
\rightarrow V~\bG~V^{-1}$, we find 
$$  (\bG \bp)_a \rightarrow V_{ab} (\bG \bp)_b ~~~~~~~~~~~(Gp)_a \rightarrow \bV _{ab}
(Gp)_b \eqno(3.3)
$$  The electric fields of Eq.(2.34) and expressions (2.37, 2.42) for $T$  are thus seen
to be invariant.

Any regulator we choose must preserve this ``holomorphic invariance". We shall use a
version of point-splitting which preserves this invariance. Explicitly we take
$$\eqalign{  p_a (\vx) \rightarrow (p_{\rm reg})_a (\vx)= \int_y \s (\vx,\vy;\e) \bigl(
K^{-1} (y,\bx) K(y,\by) \bigr) _{ab} p_b (\vy) \cr
\bp_a (\vx) \rightarrow (\bp_{\rm reg})_a (\vx)= \int_y \s (\vx,\vy;\e) \bigl( K (x,\by)
K^{-1}(y,\by) \bigr) _{ab} \bp_b (\vy) \cr}
\eqno(3.4)
$$  where
$$ 
\s(\vx,\vy;\e) ={{e^{-|\vx-\vy|^2/\e}} \over {\pi \e}} \eqno(3.5)
$$
$\e$ is the regulator parameter. As $\e \rightarrow 0,~ \s (\vx,\vy;\e) \rightarrow \d
(\vx-\vy)$ and we obtain $p,~\bp$ from the integrals in Eq.(3.4).  The factors $K^{-1}
(y,\bx) K(y,\by)$ and  $K (x,y) K^{-1}(y,\by)$ are needed so that $p_{\rm reg} (\vx)$
transforms as $\bV (\bx) p_{\rm reg} (\vx)$ and 
$\bp_{\rm reg} (\vx)$ transforms as $V (x) \bp_{\rm reg} (\vx)$.  An expression like
$K(x,\by) K^{-1} (y, \by)$ may be interpreted by the power series expansion $\sum
{{(x-y)^n} \over n!} (\del ^n K K^{-1}) (\vy)$. All terms in this expansion can be
expressed in terms of the current $J$ and its derivatives. (Similar statements hold for
$K^{-1} (y,\bx) K(y,\by)$). The regularization is thus purely a function of the currents.
The regularization (3.4) leads to 
$$\eqalign{
\bG \bp _a \rightarrow (\bG \bp _a)_{\rm reg} = \int_y \bar{\G} _{ab} (\vx,\vy) \bp _b
(\vy) \cr G p _a \rightarrow (G p _a)_{\rm reg} = \int_y \G_{ab} (\vx,\vy) p _b (\vy)
\cr}
\eqno(3.6)
$$ where
$$\eqalign{
\bar{\G} (\vx,\vy) = \int_u \bG (\vx,\vu) \s (\vu,\vy;\e) K(u,\by) K^{-1} (y,\by) \cr
\G (\vx,\vy) = \int_u G (\vx,\vu) \s (\vu,\vy;\e) K^{-1}(y,\bu) K (y,\by) \cr} \eqno(3.7)
$$  Under ``holomorphic" transformations $\G,~ \bar{\G}$ transform the same way as $G,
\bG$, i.e., $\G \rightarrow \bV \G \bV ^{-1}$ and $ \bar{\G} \rightarrow V \bar{\G}
V^{-1}$. 

Expanding the $K$'s and carrying out the integrations in Eq.(3.7) we find
$$\eqalign{
\bar{\G} _{ma} (\vx,\vy)  & =  \bG (\vx,\vy) [ \d _{ma} - e^{-|\vx-\vy|^2/\e} \bigl(
K(x,\by) K^{-1} (y, \by) \bigr) _{ma}] \cr
\G _{ma} (\vx,\vy)  & =  G (\vx,\vy) [ \d _{ma} - e^{-|\vx-\vy|^2/\e} \bigl(
K^{-1}(y,\bx) K (y, \by) \bigr) _{ma}] \cr} \eqno(3.8)
$$ As $\e \rightarrow 0$, for finite $|\vx-\vy|$, $\G,~\bar{\G} \rightarrow G,~\bG$.
\vskip .1in
\noindent{\bf a) Calculation of $\Tr [T^a \bD ^{-1} (\vy,\vx)]$}
\vskip .1in Since $\bD = \bdel +\bA = \bdel + M^{\dag -1} \bdel M^{\dag}$, we have $\bD
^{-1} (\vy,\vx) = M^{\dag -1} (\vy) \bG (\vy,\vx) M^{\dag} (\vx)$. Replacing $\bG$ by
$\bar{\G}$ we have
$$
\bD ^{-1} (\vy,\vx) _{\rm reg} = M^{\dag -1} (\vy) \bar{\G} (\vy,\vx) M^{\dag} (\vx)
\eqno(3.9)
$$ As $\vy \rightarrow \vx$, we see by power series expansion,
$$\eqalign{
 & \bar{\G} (\vy,\vx) =\bG (\vy,\vx) (1- e^{-|\vx-\vy|^2/\e}) - e^{-|\vx-\vy|^2/\e}
{{(\del K K^{-1}) (\vx)} \over \pi} +... \cr & \bar{\G} (\vx,\vx) = -  {{(\del K K^{-1})
(\vx)}
\over \pi}  \cr}
\eqno(3.10)
$$ Thus
$$\eqalign{
\bD ^{-1} (\vx,\vx) _{\rm reg} & =  -{1 \over \pi} M^{\dag -1} (\vx) (\del K K^{-1})
M^{\dag} (\vx) \cr & = -{1 \over \pi} (M^{\dag -1} \del M^{\dag} + \del M M^{-1}) (\vx)
\cr & = {1 \over \pi} (A- M^{\dag -1} \del M^{\dag}) (\vx) \cr} \eqno(3.11)
$$  This leads to Eq.(2.23).
\vskip .1in
\noindent{\bf b) $p,~\bp$ as adjoints for $d\mu (H)$}
\vskip .1in We have the representation
$$
\bp _a = -ir^{* -1} _{ak} {\delta \over {\d \vf _k}}~~~~~~~~~p _a = -ir^{ -1} _{ak}
{\delta \over {\d \vf _k}} \eqno(3.12)
$$  By direct partial integration we find
$$\eqalignno{  & \int d\mu (H)~ \overline{ p_a \psi _1} \psi _2 = \int  d\mu (H)~
\psi _1 ^* \bar{ p_a}  \psi _2 + \int [d\vf]~
 \O _a \psi _1^* \psi _2 &(3.13a) \cr & \O _a = -i {\d \over {\d \vf ^k (\vx)}} [ r^{*
-1} _{ak} (\vx) \det r] &(3.13b)
\cr}
$$  {}From the definition $\d H H^{-1} = \d \vf ^a r^* _{ab} t^b$, we have
$$  r^{* -1} _{ak} (\vx) {{\d r^{* -1}_{bl} (\vy)} \over {\d \vf ^k (\vx)}} - r^{* -1}
_{bk} (\vy) {{\d r^{* -1}_{al} (\vx)} \over {\d \vf ^k (\vy)}} = -i f^{abc} r^{* -1}
_{cl} (\vx) \d (\vx -\vy) \eqno(3.14)
$$  We want to multiply this by $r^*_{la} (\vx)$ and take $\vy \rightarrow \vx$.
Regularizing as discussed before we get, with $\Sigma _{sa} (\vy,\vx) = \s (\vy,\vx;\e)
(K(y,\bx) K^{-1} (x,\bx))_{sa}$,
$$ 
\eqalign{
\int_x r^*_{ls}(\vy) \Sigma _{sa} (\vy,\vx) r^{* -1} _{ak} (\vx) {{\d r^{* -1}_{bl}
(\vy)} \over {\d \vf ^k (\vx)}}  -\int_x r^*_{ls}(\vy) \Sigma _{sa} (\vy,\vx) r^{* -1}
_{bk} (\vy) {{\d r^{* -1}_{al} (\vx)} \over {\d \vf ^k (\vy)}} \cr = -i f^{abc} r^{* -1}
_{cl} (\vy)  r^* _{ls} (\vy) \Sigma _{sa} (\vy,\vy) \cr}
\eqno(3.15)
$$  The right hand side is seen to be zero. The first term on the left hand side is the
regularized meaning of $(\d /{\d \vf ^k (\vy)})
 r^{* -1}_{bk} (\vy) $, viz.,
$$
\int_x r^*_{ls}(\vy) \Sigma _{sa} (\vy,\vx) r^{* -1} _{ak} (\vx) {{\d r^{* -1}_{bl}
(\vy)} \over {\d \vf ^k (\vx)}}  =  \left[ {{\d r^{* -1}_{bk} (\vy)} \over {\d \vf ^k
(\vy)}} \right] _{\rm reg} \eqno(3.16)
$$  The second term can be written as
$$\eqalignno{
 -\int_x r^*_{ls}(\vy) \Sigma _{sa} (\vy,\vx) r^{* -1} _{bk} (\vy) {{\d r^{* -1}_{al}
(\vx)} \over {\d \vf ^k (\vy)}}  & =-i \int_x r^*_{ls}(\vy) \Sigma _{sa} (\vy,\vx)  \bp
_b (\vy) r^{* -1}_{al} (\vx)  \cr & = -i \Tr  \bigl[ (\bp _b (\vy) r^{* -1} _{al} )
r^{*} _{la} \bigr] _{\rm reg} &(3.17a) \cr & = i \bp _b (\vy) (\log \det r^*)_{\rm reg}
&(3.17b) \cr}
$$  Eq.(3.17a) gives the regularized meaning of $(\log \det r^*)$. Eq.(3.15) can now be
written as
$$
\bigl( {{\d r^{* -1}_{bk} (\vy)} \over {\d \vf ^k (\vy)}} \bigr) _{\rm reg} + i \bp _b
(\vy) (\log \det r^*)_{\rm reg} =0 \eqno(3.18)
$$ This tells us that $\O _a =0$ and hence that $p_a,~\bp _a$ are adjoints of each other
with the Haar measure for $H$.

There is another way to obtain the result. The regularized meaning of the measure
$d\mu(H)$ is implicitly given by the formulae for correlators such as
$<J_a(\vx) J_b(\vy)>$. For example,
$$ <J_a (\vx) J_b(\vy)> = N \int d\mu(H) e^{2c_A \S} J_a(\vx) J_b(\vy)  \eqno(3.19)
$$ where $N^{-1}= \int d\mu(H) e^{2c_A \S}$. {}From the Polyakov-Wiegmann formula,
Eq.(3.2),
$$
\bp _a (\vx) (2 c_A \S (H) ) = -i \bdel J_a (\vx) \eqno(3.20)
$$  Thus we can write
$$\eqalignno{  <J_a (\vx) J_b(\vy)> & = i \int d\mu(H) J_a(\vx) \bG (\vy,\vz) \bigl( \bp
_b (\vz) e^{2c_A \S} \bigr) &{} \cr & = -i\int  d\mu (H) e^{2c_A \S}  \bG(\vy,\vz) \bigl(
\bp _b (\vz) J_a (\vx) \bigr)  &(3.21a) \cr & = -<{\cal D}_x^{ab} \bG(\vy,\vx)> = -
{{c_A \d _{ab}} \over \pi} \del _x \bG (\vy,\vx) &(3.21b) \cr}
$$  where we assumed $p_a=\bp _a ^{\dag}$ on going to Eq.(3.21a) and used  the result
$\bp _b (\vz) J_a (\vx)= -i{\cal D}_x^{ab} \d (\vx -\vz) = -i [ {c_A \over \pi} \del _x
\d _{ab} + i f^{abc} J_c (\vx)] \d (\vx-\vz)$. The result (3.21b) agrees with the
standard conformal field theory result for the correlator $<J_a (\vx) J_b(\vy)>$, which,
for example, may be evaluated independently by operator product expansions. This
consistency requires $p_a = \bp ^{\dag}_a$ for the Haar measure.
\vskip .1in
\noindent {\bf c) K\"ahler property, self-adjointness of $T$}
\vskip .1in The regularized form of the Laplacian can be written as $\D =\D _1 + \D _2$
where
$$
 -\D _1 = - \int e^{-\F} {\d \over {\d \bar{\theta} _p (\vu)}} r^{* -1} _{rp} (\vu)
\bar{\G} _{ar} (\vx,\vu) K_{ab}(\vx) e^{\F} \G _{bs} (\vx,\vv) r^{-1} _{sq} (\vv) {\d
\over {\d \theta _q (\vv) }} \eqno(3.22)
$$
$e^{\F} = (\det r) e^{2c_A \S (H,\e)}$ where $\S (H,\e)$ is the WZW-action plus possibly
$\O (\e)$-terms and $\D _2$ is given by $\D _1$ with $
\theta,~\bar{\theta}$ exchanged. By using Eq.(3.18) we can rewrite this as
$$
 -\D _1 = - \int e^{-2c_A \S (H,\e)} \bp _r (\vu) \bar{\G} _{ar} (\vx,\vu) K_{ab}(\vx)
e^{2 c_A \S (H,\e)} \G _{bs} (\vx,\vv) p_s (\vv) \eqno(3.23)
$$  We want to move $\bp _r (\vu)$ to the right end. In moving  $\bp _r (\vu)$ through
$\bar{\G} _{ar} (\vx,\vu)$ we encounter the potentially singular term 
$\bigl[ ~\bp _r (\vu),~ \bigl(K(x,\bu) K^{-1} (u,\bu) \bigr)_{ar} \bigr]$. By writing
this as 
$$
\bigl[ ~\bp _r (\vu),~ \bigr( K(x,\bu) K^{-1} (\vu) \bigl) _{ar}\bigr] _{\rm reg}  =
\int_z
\bigl( K(u,\bz) K^{-1} (\vz) \bigr) _{rs} \s (\vu,\vz;\e)
\bigl[ \bp _s (\vz), \bigl( K(x,\bu) K^{-1} (\vu) \bigr) _{ar} \bigr] 
\eqno(3.24)
$$  and evaluating the commutator, we see that this is indeed zero. The vanishing of at
least part of this expression may be seen from the Gauss law. On gauge-invariant
functions we have $\bp _a = K_{ab} p_b$; this is essentially the Gauss law condition on
wavefunctions. Taking conjugates and writing $\bp _a ^{\dag} = p_a$ we get $p_a =\bp _b
K_{ba}$. However, directly from  $\bp _a = K_{ab} p_b$ we get $p_a = K_{ba}\bp_b$ using
$(K K^T)_{ab} = \d _{ab}$. The consistency of these expressions requires that $[\bp _b
(\vx), K_{ba} (\vx)] = 0$; the chosen regulator must give this result for consistency of
the Gauss law condition.

With this result we can write Eq.(3.23) as
$$  -\D _1 = - \int e^{-2c_A \S (H,\e)}  (\bar{\G}_{ar} \bp _r)  K_{ab} e^{2 c_A
\S (H,\e)} (\G _{bs}  p_s) \eqno(3.25)
$$  The K\"ahler property and the equivalence of the regularized form of Eq.(2.37) and
Eq.(2.42) follow if $Q=0$ where
$$
 Q= \int e^{-2c_A \S (H,\e)}  \bar{\G}_{ar} (\vx,\vu) \bigl[ \bp _r (\vu),  K_{ab}
(\vx)e^{2 c_A \S (H,\e)} \G _{bs} (\vx,\vv) \bigr]  p_s(\vv) \eqno(3.26)
$$  Using the expansion of $\bar{\G}(\vx,\vu)$ as in Eq.(3.10), it is easy to see that
the $\e$-independent part of 
$e^{2 c_A \S (H)}\bar{\G}_{ar}  [\bp _r ,  K_{ab}] $ cancels the contribution
$\bar{\G}_{ar} K_{ab} [\bp _r ,  e^{2 c_A \S (H)} ]$.  Writing $2 c_A
\S (H,\e) = 2 c_A \S (H) + \tilde{\S}$ we then find
$$\eqalign{  Q=& \int_v \Bigl[   i \int_{x,u} {e^{-|\vx-\vu|^2/\e} \over {\pi (x-u)}} 
\bdel J_r (\vu) K_{ab}(\vx) K_{an} (x,\bu) K_{rn}(u,\bu)  F_{bs} (\vx,\vv) \cr -&
\int_x  {e^{-2|\vx-\vv|^2/\e} \over {\pi (x-v)}} f_{mrl} K_{ab}(\vx) K_{an}(x,\bv)
K_{rn}(\vv) K_{lb}(\bx,v) K_{ms}(v,\bv) \cr + & \int_{x,u} \bar{\G}_{ar}(\vx,\vu)
K_{ab}(\vx) ~\bigl[ \bp _r (\vu), \tilde{S} \bigr]~ F_{bs}(\vx,\vv) \Bigr]  G(\vx,\vv)
p_s(\vv) \cr}
\eqno(3.27)
$$  where $F_{bs} (\vx,\vv) = \d _{bs} - e^{-|\vx-\vv|^2/\e} \bigl( K^{-1} (\bx,v)
K(v,\bv) \bigr) _{bs}$. For the first term in $Q$, because of the exponential
$e^{-|\vx-\vu|^2/\e} $, the contribution to the integral for small $\e$ comes from
$|\vx-\vu| {\buildrel < \over \sim}\sqrt {\e}$ and we can expand in  powers of $(u-x)$.
Likewise, for the second term, we can expand the product of the $K$'s around $v$.  In
this case we find
$$\eqalign{  Q=& -\e \int i \bigl[  \del ^2 \bar{J} _b (\vx) F_{bs} (\vx,\vv) + \12 \d
_{bs} \del ^2 \bar{J} _b (\vx) \bigr] G(\vx,\vv) p_s(\vv)  \cr & + \int \bar{\G} _{ar}
(\vx,\vu) K_{ab} (\vx) [\bp _r (\vu), \tilde{\S}] \G _{bs} (\vx,\vv) p_s(\vv) \cr}
\eqno(3.28)
$$  It is consistent to set $\tilde{\S}=0$, and we find that $Q=0$ upto $\O (\e)$-terms.
It is also possible to choose $\tilde{\S}$ of order $\e$ ( which is consistent with our
evaluation of the volume element $d\mu(\C)$) so that $Q=0 +
\O (\e ^2)$. In fact such choice is given by
$$
\tilde{\S} = -\e ~{3\pi \over {4 c_A}} \int \del \bar{J} _a  \del
\bar{J} _a + \O (\e ^2) \eqno(3.29)
$$ In any case, this checks the identity of Eq.(2.37) and Eq.(2.42).  (A similar result
can be shown directly on $\A$ without restricting to $\C$; in other words the K\"ahler
property $\del _{\ba} (g^{\ba a}g) =0$ is obtained on $\A$.) The regularized version of
Eq.(2.37) is given by
$$  T= {e^2 \over 2} \int \Pi_{rs} (\vu,\vv) \bp _r (\vu) p_s (\vv) \eqno(3.30)
$$  where
$$ 
\Pi_{rs} (\vu,\vv) = \int_x \bar{\G} _{ar} (\vx,\vu) K_{ab}(\vx) \G _{bs} (\vx,\vv) 
\eqno(3.31)
$$  Notice also that from the above calculation $\D _1 = \D _2$. The property $Q=0$ is
equivalent to a check of the self-adjointness of the expression (3.30) for $T$.
\vskip .1in
\noindent{\bf d) Checking equations of motion}
\vskip .1in The original Yang-Mills equations, in $A_0 =0$ gauge, are
$$\eqalignno{ & \dot{A} = i [T,~A] = E &(3.32a) \cr & \dot{E} = i[\H,~E]=i[V,~E]
&(3.32b) \cr}
$$ There is no contribution from $T$ in Eq.(3.32b), i.e.,
$$ [T,~[T,~A]]=0 \eqno(3.33)
$$ This equation is straightforward in $\A$, in terms of the original gauge variables.
However, it is highly nontrivial in terms of the matrix parametrization of the theory.
Its validity provides an indirect check of the consistency of the matrix reformulation
of the theory and the corresponding ordering and regularization procedures. 

{}From Eqs.(3.30, 3.31) and the fact that
$$\eqalignno{ [p_s(\vv),~ A_l(\vz)] & = - M_{ls}(\vz) \del _z \d (\vv-\vz) &(3.34a) \cr 
[\bp_r (\vu),~ A_l(\vz)] & = 0 &(3.34b) \cr}
$$  we find that
$$  [T,~ A_l(\vz)]= \int _u C_{rl} (\vu,\vz) \bp _r (\vu)
$$
$$C_{rl} (\vu,\vz) = M_{ls}(\vz) \int _x \bar{\G} _{ar} (\vx,\vu) K_{ab} (\vx) \s
(\vx,\vz;\e) (1 + {\e \over {\bx -\bz}} \del _z) [K^{-1} (\bx, z) K(z,\bz)]_{bs}
\eqno (3.35)
$$ If we take the $\e \rightarrow 0$ limit of Eq.(3.35) we find, as expected, that
$$  {\rm lim}_{\e \rightarrow 0} [T,~ A_l (\vz)] = \int_u \bG (\vz,\vu) M^{\dag}_{rl}
(\vz) \bp _r (\vu) = -i E_l (\vz)
\eqno(3.36)
$$  The evaluation of $[T,~[T,~A_l (\vz)]]$ produces three kind of terms.
$$
\eqalignno{ [T,~[T,~A_l (\vz)]] & = \int _{\o, u,v}  \Pi_{mn} (\vec{\o}, \vv) \bigl[ \bp
_m (\vec{\o}),~[p_n (\vv),~ C_{rl} (\vu,\vz)] \bigr] \bp _r (\vu) &(3.37a) \cr  & + \int
_{\o,u,v} \Pi_{mn} (\vec{\o},\vv) [p_n (\vv),~ C_{rl} (\vu,\vz)] \bp _m (\vec{\o}) \bp
_r (\vu) &(3.37b)
\cr
 & +  \Bigl( \int _{\o,u,v} \Pi_{mn} (\vec{\o},\vv) [\bp _m (\vec{\o}),~C_{rl} (\vu,\vz)]
-f_{mkr} \int_{u,v} \Pi_{mn} (\vu,\vv) C_{kl} (\vu,\vz)  & \cr  & + \int_{\o,u,v} C_{kl}
(\vec{\o},\vz) [\Pi_{rn} (\vu,\vv), \bp _k (\vec{\o})] \Bigr) ~ \bp _r (\vec{\o}) p_n
(\vv) &(3.37c) 
\cr}
$$  Evaluation of the coefficients of $\bp, \bp \bp, \bp p$ terms are quite tedious; we
eventually find that in the $\e \rightarrow 0$ limit they vanish, thus confirming
Eq.(3.33).
\vskip .1in
\noindent{\bf 4. An expression for $T$ in terms of currents}
\vskip .1in We have obtained a regularized construction of $T$ as an operator on
functions on $\C$. In this section we shall obtain an expression for $T$ in terms of
currents which can be useful in evaluating the action of $T$ on wavefunctions, which are
functions of currents.

Using expressions (3.30, 3.31) for $T$  and the chain rule of differentiation, we can
obtain the action of $T$ on a function of the currents as 
$$ T~\Psi (J)= m \left[ \int_z \omega_a(\vz){\delta \over \delta J_a(\vz)}~+\int_{z,w}
\Omega_{ab}(\vz,\vw) {\delta \over \delta J_a(\vz) }{\delta \over \delta J_b(\vw)}\right]
\Psi(J) \eqno(4.1)
$$
$$\eqalignno{
\omega_a(\vz)&= -i f_{arm} \bigl[ \del _z \Pi_{rs} (\vu,\vz) \bigr]_{\vu \rightarrow \vz}
~K^{-1}_{sm} (\vz) &{}\cr & = if_{arm} \Lambda _{rm} (\vu,\vz)  {\big |}_{\vu
\rightarrow \vz} &(4.2a) \cr
\Omega_{ab}(\vz,\vw)&= -\left[ \left[{c_A\over \pi}\partial_w \delta_{br}  +if_{brm}J_m
(\vw)\right] ~\partial_z \Pi_{rs} (\vw,\vz) \right] K^{-1}_{sa} (\vz)&{}\cr  &=
{\cal{D}} _{w~br} \Lambda_{ra} (\vw,\vz) &(4.2b) \cr}
$$  where 
$$\eqalignno{
\Lambda _{ra} (\vw,\vz) & = -(\del _z \Pi_{rs} (\vw,\vz)) K^{-1}_{sa} (\vz) &(4.3a) \cr
{\cal{D}}_{w~ab} & = {c_A\over \pi}\partial_w \delta_{ab} +if_{abc}J_c (\vw) &(4.3b) \cr}
$$ We have also used the commutation rules 
$$\eqalign{ [p_s (\vv),~J_a (\vz)] & = -i {c_A \over \pi} K_{as} (\vz) \del _z \d
(\vz,\vv) \cr [\bp _r (\vu),~ J_b (\vw)] & = -i ({\cal{D}} _w) _{br} \d (\vw-\vu) \cr}
\eqno(4.4)
$$  {}From the definition of $\Pi_{rs} (\vu,\vv)$ we find
$$\eqalign{
\Lambda _{ra} (\vw,\vz) = \int _x \bar{\G} _{mr} (\vx,\vw) G(\vx,\vz)
e^{-|\vx-\vz|^2/\e} \Bigl[ & {{\bx - \bz} \over \e} K (x, \bx) K^{-1} (z,\bx) \cr & +
K(x,\bx) \del_z ( K^{-1} (z,\bx)K(z,\bz)) K^{-1}(z,\bz) \Bigr] _{ma} \cr}
\eqno(4.5)
$$ For $\o _a(\vz)$, we need the $\vw \rightarrow \vz$ limit of $\Lambda$. The
exponential $e^{-|\vx-\vz|^2/\e}$ assures us that the contribution to the
$x$-integral is mostly from the region $|\vx-\vz|^2 {\buildrel < \over \sim} ~\e$.
Expanding around $z$, we then find
$$
\o _a(\vz) = J_a (\vz) + \O (\e) \eqno(4.6)
$$ Since $\bar{\G}$ has two terms, the expression (4.5) for $\Lambda$ splits into four
terms. 
$$\eqalignno{
\Lambda & = I +II +III + IV &(4.7a) \cr I & = {1 \over \pi} \int_x K (\vx) K^{-1} (z,
\bx) {{\s (\vx,\vz;\e)} \over {x-w}} &(4.7b) \cr II & = {\e \over \pi} \int_x K(\vx)
\del _{z} \left(  K^{-1}(z,\bx) K(\vz) \right)  K^{-1}(\vz) {{\s (\vx,\vz;\e)} \over
{(\bx - \bz) (x-w)}} &(4.7c) \cr III & =-{1 \over \pi} K(\vw) \int_x K^{-1} (x, \bw)
K(\vx) K^{-1} (z,\bx) {{\s (\vx,\vz;\e)} \over {x-w}} e^{-|\vx-\vw|^2/\e} &(4.7d) \cr IV
& = -{\e \over \pi} K(\vw) \int_x K^{-1} (x, \bw) K(\vx) \del_z \left( K^{-1} (z,\bx)
K(\vz) \right) K^{-1} (\vz) {{\s (\vx,\vz;\e) e^{-|\vx-\vw|^2/\e}} \over {(\bx
-\bz)(x-w)}}  &{}\cr &&(4.7e) \cr}
$$  We can write
$$
\s (\vx,\vz;\e) =\s (\vx,\vw;\e) ~\exp {{{(x-w)(\bz -\bw) + (\bx -\bw) (z-w)} \over \e}}
~
\exp{-{{(z-w)(\bz -\bw)} \over \e}}
\eqno(4.8)
$$  Expanding the integrands in powers of $(x-w),~(\bx -\bw)$ and performing the
$x$-integration we derive a systematic $\e$-expansion for the expressions (4.7). The
calculation is straightforward and we find 
$$\eqalignno{
 I  & =  {1\over {\pi (z-w)}}\bigl[ 1 - K(\vw)K^{-1}(z,\bw )e^{-\alpha}
\bigr]&{}\cr &- {\e \over \pi} \bigl[  {1 \over {z-w}} \bdel _z
\bigl( K(\vz) \del _z K^{-1} (\vz) \bigr) - {1 \over {(z-w)^2}} \bdel _w \bigl( K(\vw)
K^{-1} (z,\bw) \bigr) e^{-\a} \bigr] + \O (\e ^2) &(4.9a) \cr
 II  & =  {\e \over {\pi (z-w)}} \bigl[  \bigl( K \del  (\bdel  K^{-1} K) K^{-1}  \bigr)
(\vz) + K (\vw) \del _z (K^{-1} (z,\bw) K(\vz)) K^{-1}(\vz)  {{e^{-\a}}
\over { \bz -\bw}} \bigr]  &{} \cr & + \O (\e ^2) &(4.9b) \cr
 III  & =  {-1 \over {\pi (z-w)}} \bigl[  K(\vw) K^{-1} ( u, \bw )K (\vu) K^{-1} (z,
\bu) e^{- \a /2}  - K(\vw) K^{-1} (z, \bw) e^{-\a} \bigr] &{}  \cr & - {\e \over
\pi}  \bigl[ \bdel _w ( K(\vw) K^{-1} (z, \bw)) {{e^{-\a}} \over {(z-w)^2}} &{}
\cr & + 2 \del _z { 1 \over {z-w}} \bdel _z \bigl( K(\vw) K^{-1} ( u, \bw )K (\vu)
K^{-1} (z, \bu) \bigr) e^{-\a /2} & \cr & - 2 { 1 \over {z-w}} \bdel _z \bigl( K(\vw)
K^{-1} ( u, \bw )K (\vu) \del _z K^{-1} (z, \bu) e^{- \a /2} \bigr]  + \O (\e ^2)
&(4.9c) \cr
 IV & =  {\e \over {\pi (z-w) (\bz -\bw)}} \bigl[ 2 K(\vw) K^{-1} (u,\bw) K(u,\bu)
\del _z \bigl( K^{-1} (z, \bu) K(\vz) \bigr) K^{-1} (\vz)  e^{- \a /2} & \cr & - K(\vw)
\del _z \bigl( K^{-1} (z, \bw) K(\vz) \bigr) K^{-1} (\vz) e^{-\a } \bigr]  +
\O (\e ^2) &(4.9d) \cr}
$$  where $u= \12 (z+w),~\bu = \12 (\bz + \bw)$ and 
$\a = (z-w) (\bz -\bw) / \e$. Putting everything together we then find
$$ \eqalign{
\Lambda _{ra} (\vw,\vz) = & {1 \over {\pi (z-w)}} \bigl[ \d _{ra} - \bigl( K(\vw) K^{-1}
( u, \bw )K (\vu) K^{-1} (z, \bu) \bigr) _{ra} e^{- \a /2} 
\bigr] \cr & + {\e \over \pi} e^{- \a /2} \bigl[  -2 \del _z { 1 \over {z-w}}
\bdel _z \bigl( K(\vw) K^{-1} ( u, \bw )K (\vu) K^{-1} (z, \bu) \bigr)   \cr & +  { 2
\over {z-w}} \bdel _z \bigl( K(\vw) K^{-1} ( u, \bw )K (\vu) \del _z K^{-1} (z, \bu)
\bigr)     \cr & + { 2 \over {(z-w) (\bz -\bw)}}   K(\vw) K^{-1} (u,\bw) K(\vu) \del _z
\bigl( K^{-1} (z, \bu) K(\vz) \bigr) K^{-1} (\vz)    \bigr]  _{ra} \cr & + \O (\e ^2) 
\cr } \eqno(4.10)
$$  We can further expand the functions with arguments $u= \12 (z+w ), ~ \bu =
\12 (\bz + \bw)$  around ($z, ~ \bz$) to obtain
$$ \eqalign{
\Lambda _{ra} (\vw,\vz) = & {1 \over {\pi (z-w)}} \bigl[ \d _{ra} - \bigl( K(\vw) K^{-1}
( z, \bw )\bigr) _{ra} e^{- \a /2}  \bigr]  \cr +& ( {\rm
terms~of~higher~order~in~\e~or~(z-w),~(\bz - \bw)}) \cr
\equiv & ~\bar{\G}' _{ra} (\vz,\vw)+ ... \cr}
\eqno(4.11)
$$  Notice that $\bar{\G} '$ is the transpose of $\bar{\G}$ as defined in Eq.(3.8),
 with $\e$ replaced by $2\e$.  As
$\e \rightarrow 0$, $ \Lambda (\vw,\vz) \rightarrow \bar{G} (\vz,\vw)$.  For the action
of $T$ on products of currents at the same point one has to be careful. If we have only
terms of the form $\del ^n J(\vy) J(\vy)$ (for $n=0,1,...$) then one can check that
terms in Eq.(4.11)  other than $\bar{\G}' (\vz,\vw)$ do not contribute. We may thus write
$$
\Omega _{ab} (\vz,\vw) =  ({\cal{D}} _w \bar{\G}' (\vz,\vw))_{ba} + ...
\eqno(4.12)
$$ where the ellipsis refers to terms which do not contribute either for 
$z \ne w$, or for the action on terms like $\del ^n J(\vy) J(\vy)$. They may contribute
to the action of $T$ on a product like $ \bdel J(\vy) \bdel J(\vy)$. We will not
encounter products like $ \bdel J(\vy) \bdel J(\vy)$ since we shall point-separate
products of $\bdel J$'s. We shall however encounter terms like $\del ^n J(\vy) J(\vy)$
and for these the expression (4.12) suffices.

The kinetic energy term of Eq.(4.1) now becomes 
$$  T \Psi (J) = m \left[ \int J_a (\vz) {\d \over {\d J_a (\vz)}} + \int \bigl(
{\cal{D}} _w \bar{\G}' (\vz,\vw) \bigr) _{ab} {\d \over {\d J_a (\vw)}} {\d \over {\d
J_b (\vz)}} \right]  \Psi (J) + \O (\e) 
\eqno (4.13)
$$  In arriving at the expression (4.6) for $\o _a$, we have cancelled powers of
$(z-w)$ against $\bar{G} (\vz,\vw)$. Keeping track of these more carefully one finds
$$  T \Psi (J) = m \int _{z,w}\left[  \bdel J_a (\vw) \bar{G} (\vz,\vw) {\d \over {\d
J_a (\vz)}} +  \bigl( {\cal{D}} _w \bar{\G}' (\vz,\vw) \bigr) _{ab} {\d \over {\d J_a
(\vw)}} {\d \over {\d J_b (\vz)}} \right]  \Psi (J) + \O (\e) 
\eqno (4.14)
$$  A partial integration in the first term takes us back to Eq.(4.13). Under a
holomorphic transformation $J_a (\vz) \rightarrow V_{ab} J_b (\vz) + (c_A / \pi ) (\del
V V^{-1}) _a $ and $ {\d \over{\d J_a}} \rightarrow V_{ab} {\d \over{\d J_b}} $.
Expression (4.14) has manifest holomorphic invariance.
\vskip .1in
\noindent{\bf 5. A digression on the Abelian case}
\vskip .1in The next step in our discussion is naturally to consider eigenstates of $T$. 
However, to clarify the nature of some of the terms which arise, we shall, in this
section, consider the Abelian case with an added charge density due to matter fields.
 In the Abelian case we have $A= -\del \theta, ~ \bar{A} = \bdel \bar{\theta}$. (Recall
that we are using antihermitian components for the potentials.) Writing $\theta = \chi +
i \phi$ with $\chi,~ \phi$ real, we see that $\phi$ corresponds to the gauge part of
$A$. With this splitting
$$ E= {1 \over 4} \int \bar{G} \bigl( {\d \over {\d \chi}} + i {\d \over {\d
\phi}} \bigr) ~,~~~~~~
\bar{E}= -{1 \over 4} \int G \bigl( {\d \over {\d \chi}} - i {\d \over {\d
\phi}} \bigr)
\eqno(5.1)
$$  This gives $\del _i E_i = i {\d \over {\d \phi}}$ and the Gauss law condition for
physical states becomes
$$  (i {\d \over {\d \phi}} - \rho) \Psi = 0  \eqno(5.2)
$$  This has the solution
$$
\Psi (\theta) = e^{-i\int \rho \phi }~~ \Phi (\chi) \eqno(5.3)
$$ 

The kinetic energy operator becomes
$$  T = {e^2 \over 8} \int \Pi (\vu,\vv) \bigl( {\d \over {\d \chi (\vu)}} + i {\d \over
{\d \phi (\vu)}} \bigr)
\bigl( {\d \over {\d \chi (\vv)}} - i {\d \over {\d \phi (\vv)}} \bigr)
\eqno(5.4)
$$  where, in this Abelian limit,  $\Pi (\vu,\vv)$ is given by
$$ 
\Pi (\vx,\vy) =\int_u \bG (\vu,\vx) (1- e^{-|\vu-\vx|^2/\e}) G(\vu,\vy)(1-
e^{-|\vu-\vy|^2/ \e})
\eqno(5.5)
$$ The regulated Green's functions in the Abelian limit are given by
$$ \eqalign{
 \bar{\G} (\vx,\vy) & \equiv {1 \over {\pi \xi}} ( 1 - e^{-{ {\xi \bar{\xi}} \over
\e}}) = \int {{d^2 k} \over {(2 \pi )^2}} {{e^{i \vk \cdot \vec{\xi}} e^{- \e \vk ^2/4}} 
\over {i \bar{k}}} \cr
 \G(\vx, \vy) & \equiv {1 \over {\pi \bar{\xi}}} ( 1 - e^{-{{\xi \bar{\xi}} \over \e}})
= \int {{d^2 k} \over {(2 \pi )^2}} {{e^{i \vk \cdot \vec{\xi}}  e^{- \e \vk ^2 /4}}
\over {i k}}
\cr}\eqno(5.6)
$$ where $\xi =x-y$. Using these expressions in Eq.(5.5), we may write
$$
\Pi (\vx,\vy) = \int _u \bar{\G} (\vu,\vx) \G (\vu,\vy) = \int {{d^2 k} \over {(2 \pi
)^2}} {{e^{i \vk \cdot (\vx-\vy)} e^{- \e \vk ^2/2}} \over {k \bar{k}}} 
\eqno(5.7)
$$ This integral can be evaluated after introducing an infrared cutoff $R$ as
$$\eqalign{
\Pi (\vx,\vy) &= {1\over \pi}\left[ {\rm Ein}(s /2R^2 ) -{\rm Ein}(s /2\e ) +
\log (R^2/ \e )\right]\cr &\approx {1\over \pi}\left[ \log (R^2/\e ) - {\rm Ein} (s/2\e
)\right]\cr}
\eqno(5.8)
$$ where $s= |\vx-\vy| ^2$ and
$$ {\rm Ein} (z) = \int _0 ^z {dt\over t}(1-e^{-t} )\eqno(5.9)
$$ We have used the fact that $R^2 \gg s$, so ${\rm Ein} (s / 2R^2) \approx {\rm Ein}
(0) = 0$. As $\e \rightarrow 0$ for fixed $s$, 
${\rm Ein}(s/ 2\e ) \approx \log(s/2 \e)$ [12].

Using Eq.(5.3) we can write
$$ 
\eqalign{ T ~\Phi (\chi) &= {e^2 \over 8} \int \Pi (\vu,\vv) \bigl( {\d \over {\d
\chi(\vu)}} +
\rho (\vu) \bigr) \bigl( {\d \over {\d \chi(\vv)}} -\rho (\vv) \bigr) \Phi (\chi)\cr &=
\left[ {e^2\over 8} \int \Pi (\vu,\vv ) {\d \over {\d \chi(\vu)}}  {\d \over {\d
\chi(\vv)} }~- {e^2\over 8} \int \Pi (\vu,\vv ) \rho (\vu) \rho (\vv)\right] \Phi(\chi
)\cr}
\eqno(5.10)
$$  The term quadratic in $\rho$ is the Coulomb interaction,
$$  T_{Coul} = -{e^2 \over 8} \int \Pi (\vu,\vv) \rho(\vu) \rho (\vv)
\eqno(5.11)
$$ For a two-body state with $ \rho (\vu) = \d (\vu-\vx) - \d (\vu-\vy)$, we get
$$\eqalign{  T_{Coul} &=-{e^2 \over {8 }} \bigl[ \Pi (\vx,\vx)+ \Pi (\vy,\vy) - \Pi
(\vx,\vy) -\Pi (\vy,\vx)\bigr]\cr &\approx -{e^2\over 4\pi} \log (|\vx-\vy|^2/2\e )\cr}
\eqno(5.12)
$$  This is indeed the expected logarithmic Coulomb interaction.  However, its
dependence on the short distance cut-off $\epsilon$ deserves comment.  Going back to
expression (5.11), we see that a change of scale in $\Pi(\vu, \vv)$, say, $R \rightarrow
\alpha R$ produces a correction of the form $\log \alpha \int \rho (\vu) \int \rho
(\vv)$ in
$T$, which is zero for states with total charge zero, as for the two-body state we are
considering. The Coulomb interaction is thus expected to be independent of the cut-off
scales; some physical scale $\lambda$ should appear in the logarithmic term. The reason
why $\epsilon$ appears in Eq.(5.12) is that, with our regulator, the self-energy
subtractions are also  automatically done at scale $\epsilon$. This can be clarified by
considering two matter fields, say, $\psi$ and $\zeta$ of positive and negative unit
charges respectively. Thus $\rho = \psi^\dagger \psi ~- \zeta^\dagger \zeta$. The
two-body state of zero total charge is given by
$$ |x,y\ra = \psi^\dagger (x) \zeta^\dagger (y) ~ |0\ra
\eqno(5.13)
$$ This is an eigenstate of $\rho$ with eigenvalue $ [\d (\vu-\vx) - \d (\vu-\vy)]$ and
leads to the result (5.12). We can write the product of the charge densities as
$$
\rho (\vu) \rho (\vv) = : \rho(\vu) \rho(\vv) : ~+ \delta( \vu, \vv) \left( \psi^\dagger
\psi + \zeta^\dagger \zeta \right) \eqno(5.14)
$$ where the colons indicate normal ordering. Thus
$$ T_{Coul} = -{e^2 \over 8} \int \Pi (\vu,\vv) :\rho(\vu) \rho (\vv): ~-{e^2 \over 8} 
\Pi(\vu, \vu) \int \left( \psi^\dagger \psi + \zeta^\dagger \zeta \right)
\eqno(5.15)
$$ The second term is a correction to the mass of the matter fields. Indeed, if we have
a mass term ${\cal H}_{mass} = m  \int \left( \psi^\dagger \psi + \zeta^\dagger \zeta
\right)$, we see that the correction $-(e^2/8 )\Pi(\vu,\vu)$ can be absorbed into the
definition of mass. Alternatively, we can introduce a renormalized mass $m_{ren}$
defined at scale $\lambda$ by
$$ m = m_{ren} -{e^2\over 8\pi} \log({2\e / \lambda }) \eqno(5.16)
$$ The energy of the two-body state now becomes
$$ (T_{Coul} + {\cal H}_{mass}) |x,y\ra = \left[2 m_{ren} - {e^2\over 4\pi}  \log
(|\vx-\vy|^2/\lambda)\right]~|x,y\ra \eqno(5.17)
$$ As expected, the subtraction scale $\lambda$ appears in the Coulomb interaction.

We can also phrase this as follows. We do not need to introduce a mass term or,
equivalently, we can set $m_{ren}=0$. Instead, the properly regularized $T_{Coul}$ is
defined as
$$\eqalign{ T_{Coul}(\lambda) &=  -{e^2 \over 8} \int \Pi (\vu,\vv) \rho(\vu) \rho (\vv)
+ {e^2\over 2} \log ({2\e / \lambda}) ~{\cal Q}\cr {\cal Q}&= -{1\over 4\pi} \int \left(
\psi^\dagger \psi + \zeta^\dagger \zeta \right)\cr}
\eqno(5.18)
$$ We introduce a new operator ${\cal Q}$ which gives self-energy subtractions at the
desired scale. Obviously, for $\lambda =2 \epsilon$, viz., subtractions at scale
$\epsilon$, we go back to the expression (5.12).

This latter point of view of adding an operator ${\cal Q}$ is more appropriate for the
non-Abelian case, where the mass is dynamically generated.
\vskip .1in
\noindent{\bf 6. Construction of eigenstates of $T$}
\vskip .1in We now turn to the construction of eigenstates of $T$. The lowest eigenstate
is given by $\Psi_0 =$ constant, since $T$ involves derivatives. We may take the
normalized state as $\Psi _0 =1$ since $\int d \mu (\C) = 1 $. The state with the lowest
number of $J$'s we can construct, which also has holomorphic invariance, is
$$
\Psi _2 = \int _{x,y} f(\vx,\vy) \bigl[ \bdel J_a (\vx) \bigl( K(x,\by) K^{-1} (y,
\by) \bigr) _{ab} \bdel J_b (\vy)  \bigr] 
\eqno(6.1)
$$ The term $K(x,\by) K^{-1} (y, \by)$ ensures the holomorphic invariance of the product
of two currents in the above expression.  The term $K(x,\by) K^{-1} (y, \by)$ can also
be written in terms of currents and derivatives of currents by a Taylor expansion
$$ K(x,\by) K^{-1} (y, \by)  = \sum _0 ^{\infty} {{(x-y)^n} \over n!} \bigl( \del ^n K
K^{-1} \bigr) (y,\by)  
\eqno(6.2)
$$  The lowest order term in $\Psi _2$ has two currents (two $\bdel J$'s). $\Psi _2$ is
in general not an eigenstate of $T$; the action of $T$ can generate terms which have at
least three currents, four currents and so on. These terms generally come with powers of
$(x-y)$. By taking $(x-y)$ small we can avoid such terms and obtain an eigenstate. It is
instructive to keep the separation $(x-y)$ arbitrary for the moment and evaluate the
action of $T$ on
$\Psi_2$. We find
$$ 
\eqalign{ T \Psi_2 = 2m &  \int _{x,y} f(\vx,\vy) \Bigl\{\bdel J_c (\vx)
\bigl( K(x,\by) K^{-1} (y,
\by) \bigr) _{ab} \bdel J_b (\vy)
\bigl[ \d _{ca} +{\cal V}_{ca}(\vx,\vy)\bigr]\cr &+  {c_A {\rm dim}G \over \pi}
 \del_x \bdel_x  \s (\vx,\vy;\e) \Bigr\} ~+ \int \O \bigl( (x-y) J^3 \bigr) f(\vx,\vy)
\cr}
\eqno(6.3)
$$  where $\O \bigl( (x-y) J^3 \bigr)$ refers to terms which have at least three
currents and one power of $(x-y)$.  Also ${\cal V}(\vx,\vy)$ is defined by
$$ {\cal V}_{ca}(\vx,\vy)= {\pi \over  {2 c_A}} (T^k T^l)_{ca} \Bigl\{
\left[\Pi (x,\bx,~x,\by )- \Pi (x,\bx,~y,\by ) \right] K^{-1}(x,\by )\Bigr\}_{kl}
\eqno(6.4)
$$  where $\Pi (u,\bu,~v,\bv)=\Pi (\vu,\vv)$ is defined by Eq.(3.31). From the
transformation properties of $\Pi (\vu,\vv)$ and hence of $\V_{ca}(\vx,\vy)$, the
holomorphic covariance of Eq.(6.3) can be verified. We can think of the value $2m$ as
arising from one factor of $m$ for each $\bdel J$ in
$\Psi_2$ which is in accord with Eq.(2.44). It is like $\Psi_2$ has two constituent
particles each represented by $\bdel J$. 
${\cal V}_{ca}(\vx,\vy)$ is thus an interaction potential for the two currents.
 
For most of the terms in the above calculation of
$T \Psi_2$, the naive replacement of $\bar{\G}'$ by
$\bar{G}$ suffices. Only the terms involving ${\cal V}_{ca}$ in Eq.(6.3) require more
careful treatment. This arises from $\bdel J_a (\vx)$ $\bigl( K(x,\by) K^{-1} (y,
\by) \bigr) _{ab}$ $ \bdel J_b (\vy)$ when one of the 
${\d / {\d J}}$'s in Eq.(4.13) acts on a $\bdel J$ and the other on $K(x,\by) K^{-1} (y,
\by)$.  One can evaluate this by using the power series expansion of Eq.(6.2). A simpler
method is to use the expression for $T$ in terms $p_a, \bp_a$ as in Eq.(3.30) for this
particular term. This is what we have done and leads directly to the result of Eq.(6.3).

There are a number of interesting points to be made regarding Eq.(6.3). First of all, it
is easy to see that  the leading term in an expansion around the Abelian limit is given
by
$$\eqalign{ {\cal V}_{ca}(\vx,\vy) & \approx  \delta_{ca}~ {\pi  \over 2} \{ \Pi
(x,\bx,~x,\by) - \Pi (x,\bx,~y,\by )\}\cr &\approx   {\d_{ca} \over 2} {\rm Ein}(s/2\e )
~\approx   {\d_{ca} \over 2}
\log (s/2\e )\cr}\eqno(6.5)
$$ where we have used Eq.(5.8). Comparison with the Abelian limit shows that this is
indeed the logarithmic  Coulomb potential between the two constituent particles of
$\Psi_2$.

Consider now the $\e$-dependence of Eq.(6.5). $\e$ is a short distance  cut-off and we
should expect physical results to be independent of $\e$. $\V(\vx,\vy)$ is properly
regulated at short distances so that $\V(\vx,\vx) =0$. In analogy with the Abelian case,
we see that this corresponds to the subtraction of Coulomb self-interactions at the
scale $\e$.  In order to obtain subtractions of self-energy at some other desired scale
$\lambda$, we must introduce the operator ${\cal Q}$. For the non-Abelian theory, this
can be defined as
$$\eqalign{ {\cal Q}&= \e \int \Pi'_{rs}(\vu,\vv) p_r^\dagger (\vu) p_s(\vv)\cr
\Pi'_{rs}(\vu,\vv)&= \s (\vu,\vv;\e) K_{rs} (u, \bv) \cr}
\eqno(6.6)
$$ where $p_r^\dagger$ is the adjoint of $p_r$ including the $\exp(2c_A \S (H))$ term in
the measure of integration, i.e., $p_r^\dagger = (\bp_r -i
\bdel J_r)$.
${\cal Q}$ is a self-adjoint operator. The action of ${\cal Q}$ on $J_a$ is proportional
to
$[\partial_x \Pi'(\vu,\vx)]_{u\rightarrow x}$ which is easily checked to be zero. Thus 
adding a term proportional to ${\cal Q}$ to $T$ would not change the result of Eq.(2.44).

We now calculate the action of ${\cal Q}$ on $\Psi_2$.  Because of the prefactor
$\e$ in the definition, most of the terms in ${\cal Q} \Psi_2$  are zero, at least as
$\e \rightarrow 0$; only one term\hfil\break 
$\e \int \Pi'_{rs} [\bp_r(\vu),~\bdel J_a(\vx) ] [p_s(\vv),~
(K(x,\by)K^{-1}(y,\by))_{ab}] \bdel J_b(\vy) $  gives a nonzero contribution. We get
$$ {\cal Q} \Psi_2 = {c_A\over \pi} \Psi_2  ~+... \eqno(6.7)
$$ where the ellipsis refers to terms which vanish as $\e \rightarrow 0$; such terms are
of the order of $\e$ and still vanish  if we multiply ${\cal Q}$ by a factor
proportional to $\log \e$.  

We now define the regularized expression for $T$, with self-energy subtractions at scale
$\lambda$ as
$$ T(\lambda )= T + { e^2 \over 2} \log ({2\e / \lambda })~ {\cal Q}
\eqno(6.8)
$$ Using Eq.(6.7), the action of $T(\lambda )$ on $\Psi_2$ is easily evaluated as
$$ 
\eqalign{ T(\lambda )~ \Psi_2 = 2m &  \int _{x,y} f(\vx,\vy) \Bigl\{\bdel J_c (\vx)
\bigl( K(x,\by) K^{-1} (y,
\by) \bigr) _{ab} \bdel J_b (\vy)
\bigl[ \d _{ca} +{\cal V}_{ca}(\vx,\vy) + \12 \d_{ca} \log ({2 \e / \lambda})\bigr]\cr
&+  {c_A {\rm dim}G \over \pi}
 \del_x \bdel_x  \s (\vx,\vy;\e) \Bigr\} ~+ \int \O \bigl( (x-y) J^3 \bigr) f(\vx,\vy)
\cr}
\eqno(6.9)
$$  The new potential is given by
$$
\V_{ca}(\vx,\vy) + \12 \d_{ca} \log ({2 \e / \lambda})
\approx   \12 {\d_{ca} }  \log (|\vx-\vy|^2/\lambda )\eqno(6.10)
$$ The result for the potential (at finite nonzero separation $|\vx-\vy|$) is
independent of $\e$ as expected; the limit $\e \rightarrow 0$ can now be taken without
difficulty. The scale factor $\lambda$ enters the expression for the energy.  The former
expression (6.3) is also seen to be the special case of $\lambda = 2 \e$.

${\lambda}$ is a physical scale parameter. However, since $T/m$ is a scale-invariant
operator, the numerical value of ${\lambda}$ cannot be determined by consideration of
$T$ alone; it can be freely chosen as far as eigenstates of $T$ alone are concerned. The
inclusion of the potential energy term will determine what $\lambda$ should be; we
expect it to be of the order of $(1/m^2)$ itself.

Generally speaking,  $\Psi_2$ cannot be an eigenstate because of the Coulomb-like
interaction and because of  $\O ((x-y) J^3)$ terms. However, if we are only interested
in constructing an eigenstate of $T$,  we can use an appropriate $f(\vx,\vy)$ which
gives a specific value  to the interaction energy and take a limit where the terms $\O
((x-y) J^3)$ in Eq.(6.9) can be neglected.  We do this by first taking 
$|\vx-\vy| \approx {\sqrt {\lambda '}}$, which can be achieved by choosing $f(\vx,\vy)$
to be 
$$  f(\vx,\vy) = {{e^{-|\vx-\vy|^2 / \lambda '}} \over { \pi \lambda '}} f(\vec{X}) = 
\s (\vx,\vy; \lambda ' ) f(\vec{X}) \eqno(6.11)
$$  where $\vec{X} = \12 (\vx+\vy)$ is the center of momentum  coordinate. Using the
above form of $f(\vx,\vy)$ and carrying out  the integration over the relative
coordinate, we find  the leading  term of $T(\lambda )\Psi_2$ in the Abelian limit to be
$$\eqalign{ T(\lambda )\Psi_2 &= \int \s (x,y;\lambda' )~ 2m \left( 1+\12 {\rm Ein}
(s/2\e ) +\12
\log (2 \e / \lambda ) \right) f(\vec{X})\cr &~~~~~~~~~~~~ \bigl[ \bdel J_a (\vx) \bigl(
K(x,\by) K^{-1} (y,
\by) \bigr) _{ab} \bdel J_b (\vy)  \bigr] +...\cr &= 2m \left( 1+ \12 \log (\lambda' /
\lambda ) \right) \Psi_2\cr}
\eqno(6.12)
$$ (The term involving $(c_A {\rm dim G}/\pi ) \partial\bdel \s (\vx,\vy; \e)$ is not
included in this expression; this term is discussed below.)

In order to eliminate the ${\cal O}((x-y)J^3)$ terms in Eq.(6.9) which are of the order
of $\lambda'$, we shall take $\lambda'$ very small.  As we have  already mentioned,
$\lambda$ is not determined by $T(\lambda )$ alone. For obtaining an eigenstate of
$T(\lambda )$, we may thus take $\lambda'$ small, but with a fixed value for $(\lambda'
/\lambda)$. (Of course, we must also have $\lambda , \lambda' \gg \e$.) As will be clear
from the next section, the perturbative inclusion of the  potential energy term is valid
only for the low momentum modes,
$\lambda $ giving the scale for the distinction between low and high  momentum. Thus for
consistency, we must also have 
$\lambda' {\buildrel > \over \sim}\lambda$. Again, as far as $T$ alone is concerned, 
the numerical value of the ratio $(\lambda' /  \lambda )$ is undetermined; $\lambda$
will be fixed by inclusion of the potential energy term in ${\cal H}$ and $\lambda'$
will be determined by balance of kinetic and potential terms via the uncertainty
principle or equivalently by solving a Schrodinger-like equation.

The remaining term 
$$
\int {c_A {\rm dim}G \over \pi}f(\vx,\vy) \del_x\bdel_x \sigma(\vx,\vy;\e ) = - \int_X
{c_A {\rm dim}G\over {\pi ^2 \lambda  ^{'2} }}f(\vec{X})\eqno(6.13)
$$ is a constant normal-ordering correction for 
$\bdel J \bdel J$. Combining the above equation with Eq.(6.12) we see that the state
$$\eqalign{
\tilde{\Psi} _2 & = \int_X f(\vec{X}) \Bigl[ \int _{\xi} \s (\vx,\vy;\lambda ' ) \bdel J
_a (\vx)
\bigl( K(x,\by) K^{-1} (y,\by) \bigr) _{ab} \bdel J_b (\vy) - { {c_A {\rm dimG}} \over {
\pi ^2 \lambda  ^{'2}}}  {1\over {(1+\12 \log ({\lambda' / \lambda}) }}\Bigr] \cr  &
\equiv \int_X f(\vec{X}) : \bdel J \bdel J : (\vec{X}) \cr}
\eqno (6.14a)
$$  is an eigenstate of $T$ with eigenvalue $2m (1+\log ({\lambda' / \lambda }) ) $, as
$\lambda ' \rightarrow 0$, i.e.,
$$ \eqalign{ T (\lambda )~\tilde{\Psi} _2 &= 2m_*  ~\tilde{\Psi} _2 \cr m_* &= m (1+\12
\log ({\lambda' /  \lambda }) ) \cr}
\eqno (6.14b)
$$ A special choice is to take $\lambda' =\lambda$, in which case, it is easily checked
by direct computation that $\tilde \Psi_2$ is orthogonal to the ground state we have
obtained. In the limit of very small $\lambda ,~\lambda'$ with $(\lambda ' / \lambda )=1$
we clearly have  an excited eigenstate of $T(\lambda )$ with eigenvalue $2m$.

Since $\tilde{\Psi} _2$ is a function of the currents, the normalization presents no
difficulties. The normalization condition becomes \footnote{*}{We thank G. Alexanian for
the computation of this condition.}
$$ {{c_A ^2 ({\rm dimG})^2} \over {6 \pi ^3}} \int \bdel ^3 f ~\del ^3 f^* =1
\eqno (6.15)
$$
\vskip .1in
\noindent{\bf 7. Corrections due to the potential term}
\vskip .1in 
So far we have considered the kinetic term $T$ by itself and obtained $\Psi
_0,~\tilde{\Psi}_2$ as eigenstates of $T$.  Concerning the diagonalization and
construction of eigenstates of $T$ there are two different points of view.
Mathematically $T$ is proportional to the Laplacian on the configuration space
$\C$ and one can ask what the eigenstates are, independently of the Yang-Mills
Hamiltonian. The question of how good an approximation $T$ is to $(T+V)$ is irrelevant
for this and the discussion of section 6 is directly applicable. However if we regard
the diagonalization of $T$ as an approximation to the diagonalization of $(T+V)$, we see
that this is a meaningful starting point only for modes of momenta $k \ll m$. For, as
will be clear soon, the potential energy term gives contributions of the order $\vk
^2/m^2$ where $k$ is a typical momentum. Part of the potential term pertaining to modes
of momenta $k \ll m$ can be treated in an expansion in $1/m$. 

We write the potential term as
$$\eqalign{ V & = {\pi \over {m c_A}} \int_x : \bdel J_a (\vx) \bdel J_a  (\vx) : \cr &
= {\pi \over {m c_A}} \bigl[ \int_{x,y} \s (\vx,\vy;\lambda ) \bdel J_a (\vx) (K(x,\by)
K^{-1} (y,\by))_{ab} \bdel J_b (\vy) - {{c_A {\rm dim} G} \over {\pi^2
\lambda^{2}}} \bigr] \cr}
\eqno(7.1)
$$
Since the kinetic term has been defined with a subtraction scale $\lambda$,
we are using the same value in defining the potential term as well.
$\Psi _0=1$ is the lowest order result for the vacuum wavefunction. To include the
correction due to $V$, we consider $e^P$ where $P$ can be expanded in powers of $1/m$
with
$$ P = \b V + \O (1/m^3) \eqno(7.2)
$$ and $\b \simeq 1/m$. We find
$$\eqalign{ e^{-P} \H e^P & = e^{-P} (T+V) e^P \equiv T + [T,P] +V + \O (1/m^2)
\cr & = T + (2m \b +1) V + {{4 \pi \b} \over c_A} \int ( {\cal{D}} \bdel J) {\d \over \d
J} + \O (1/m^2) \cr}
\eqno(7.3)
$$ We have used the result (6.14). Choosing $\b = - 1/(2m)$ we find
$$\eqalign{
\H e^P ~\Psi_0 & = e^P [T - {{2 \pi} \over {m c_A}} \int {\cal{D}} \bdel J {\d \over {\d
J}} + ...] ~ \Psi _0
\cr & = 0 + \O (1/m^2) \cr}
\eqno(7.4)
$$ Thus $e^P =e^{-V/2m}$ gives the corrected vacuum wavefunction to order $1/m$.

The expectation value of an operator $\O$ in this corrected vacuum can be written as
$$ 
\la 0|\O|0\ra  = \int d\mu (\C) \exp\left[{-{V \over m}} \right]~~\O \eqno(7.5)
$$  This is the functional integral for two-dimensional (Euclidean)  YM theory of
coupling constant $g^2 = me^2 = {{e^4 c_A} \over {2\pi}}$. For the Wilson loop operator
$W_R (C)$ in the representation $R$, we can use the results of references [10] to obtain
$$ <W_R (C)> \simeq \exp\left[ - {{e^4 c_A c_R} \over {4 \pi}} A_C\right]
 \eqno(7.6)
$$ where $A_C$ is the area of the curve $C$. This result pertains only to the
contribution of modes of momenta $k\ll m$. The high-momentum modes can give a
contribution which goes like the perimeter due to the correlation of currents at nearby
points on $\C$ and this can dominate for large loops.

The action of $\H$ on a perturbed state $e^P J_a$ gives
$$
\H (e^P J_a (\vx)) \simeq (m- { \nabla ^2_x\over 2m}) (e^P J_a (\vx)) + {{2i  \pi} \over
{m c_A}} f_{abc} e^P J_b (\vx) \bdel J_c (\vx) +... \eqno(7.7)
$$  This is not quite an eigenstate; however the corrected energy starts off as
$m + \vk ^2/2m$ for momentum $k$. A $(1/m)$-expansion is necessarily a nonrelativistic
expansion and this result is just what we expect. This is similar to what happens with
solitons and one must sum up a sequence of terms to obtain the relativistic result
$\sqrt{\vk ^2 + m^2}$   [14] (see next section).

The perturbative inclusion of the potential energy applies to low momentum 
modes. For the other modes, one must seek a diagonalization of the high
momentum part of $(T+V)$, perhaps along the lines of the next section, and match with
the low momentum expansion. This matching, among other things, will determine the scale
${\lambda}$ introduced in section 6. 

Notice also that if we include the potential energy $V$ as above, the action of ${\cal
H}$ on $\Psi_2$ gives a result of the form
$$\eqalign{ {\cal H} ~ e^P \Psi_2 = e^P ~\int _{x,y}  \bdel J_a (\vx) &\bigl( K(x,\by) K^{-1} (y,
\by) \bigr) _{ab} \bdel J_b (\vy)\times \cr &\left[ 2m\bigl(1 + \12 \log [{|\vx-\vy|^2
/  \lambda }] \bigr) -{ \nabla_x^2\over 2m} -{\nabla_y^2\over 2m} \right] f(\vx, \vy)
+...\cr}
\eqno(7.8)
$$ We see that eigenstates can be constructed by taking $f(\vx,\vy)$ to be solutions of
the two-body Schrodinger equation
$$
\left[ -{ \nabla_x^2\over 2m} -{\nabla_y^2\over 2m} ~ +2m\bigl(1 + \12 \log
[{|\vx-\vy|^2 /  \lambda }] \bigr)  \right] f(\vx, \vy) ~= E f(\vx,\vy) 
\eqno(7.9)
$$ The states so obtained will be the orbital excitations of the basic two-body state.
Of course, to do this properly one must go beyond the nonrelativistic approximation and
 the lowest order logarithmic potential.
\vskip .1in
\noindent{\bf 8. A consistent truncation}
\vskip .1in Qualitatively, the emergence of the mass gap is the most interesting
nonperturbative effect. As we have argued, this has to do with the $e^{2c_A
\S (H)}$ factor in $d\mu (\C)$. A perturbation theory around the Abelian limit (which is
an expansion in powers of the structure constants $f_{abc}$) would not see this effect;
however having obtained the factor $e^{2c_A \S (H)}$, an improved perturbative expansion
can be done.

We write $H=e^{t_a \vf_a}$ and do an expansion in powers of $\vf$ for $T$,
$J_a$ and the WZW action $\S  (H)$. This is equivalent to an expansion around the
Abelian limit for these terms. For example,
$$  2c_A \S (H) \equiv -{c_A \over 2\pi} \int \del\vf _a \bdel \vf _a + ...
\eqno(8.1)
$$  We will however retain the factor $e^{2c_A \S (H)} \simeq e^{- {c_A \over 2\pi}
\int \del \vf \bdel \vf}$ rather than expanding this as $(1- {c_A \over 2\pi}
\int \del \vf \bdel \vf + ...)$.  This expansion is thus not  the same as expansion
around the Abelian limit for the full theory. To the lowest order in the $\vf$'s we find
$$
\eqalignno{ & d\mu (\C) \simeq [d \vf] e^{- {c_A \over 2\pi} \int \del \vf  _a
\bdel \vf _a} &(8.2a) \cr & T= T_1 + T_2 &(8.2b) \cr & T_1 \simeq m \int \vf _a {\d
\over \d \vf _a} ~;~~~~~~~~T_2 \simeq {m\pi \over c_A} \int C(\vx,\vy) {\d \over {\d \vf
_a (\vx)}} {\d \over {\d \vf _a (\vy)}} &(8.2c) \cr & C(\vx,\vy) = \int _z
\bar{G} (\vx,\vz) G(\vz,\vy) = - \int { d^2 k \over (2\pi)^2} { e^{i \vk \cdot
(\vx-\vy)} \over {k\bar{k}} }&(8.2d) \cr & V \simeq {c_A \over m\pi} \int \del \vf _a
(-\del
\bdel) \bdel \vf _a &(8..2e) \cr}
$$  This expansion is consistent in the sense that the self-adjointness of $T$ and $V$
is respected. It is in fact instructive to consider the self-adjointness of $T$ as given
above. We find
$$
\eqalignno{
 \la \psi_1|T_1 \psi_2\ra    =  &- \la T_1 \psi_1 | \psi_2\ra - m \d (0) \int d^2 x ~{\rm
dim} G \la \psi_1|\psi_2\ra   &{} \cr & + {{m c_A} \over \pi} \la \psi_1 | \int \bdel\vf
\del \vf | \psi_2\ra  &(8.3a) \cr
 \la \psi_1|T_2 \psi_2\ra   =&  \la T_2 \psi_1 | \psi_2\ra +  m \d (0) \int d^2 x ~{\rm
dim} G \la \psi_1|\psi_2\ra  &{} \cr
 & + 2\la T_1\psi_1|\psi_2\ra - {{m c_A} \over \pi} \la \psi_1 | \int \bdel\vf \del \vf
| \psi_2\ra  &(8.3b) \cr
  \la \psi_1|(T_1+T_2) \psi_2\ra   =  &  \la (T_1+T_2) \psi_1 | \psi_2\ra   &(8.3c) \cr}
$$  Eventhough formal expressions like $\d (0)$ and $\int d^2 x$ occur here, these
equations illustrate the main point, viz., that $T_2$ is not self-adjoint by itself;
$T_1$, which is the crucial term for the mas gap, is needed for self-adjointness so long
as we have the factor $exp\left[ -{c_A \over 2\pi} \int \del
\vf \bdel \vf\right]$ in $d\mu (\C)$. (The cancellation of formal expressions involving
$\d (0)$ and $\int d^2 x$ need not worry us at this stage; these arise from the
truncations. We have already checked that $T^{\dag} = T$ in the regulated version.)

We now absorb the factor $exp\left[ -{c_A \over 2\pi} \int \del
\vf \bdel \vf\right]$ into the wavefunctions, defining $\Phi = e^{-{c_A \over 4\pi} \int
\del \vf \bdel
\vf} \Psi$, so that
$$ 
\la 1|2\ra  \simeq \int [d\vf] ~ \Phi_1^* \Phi_2 \eqno(8.4)
$$  For the wavefunctions $\Phi$ we get, upto an additive constant,
$$
\H = {m\pi \over c_A} \int C(\vx,\vy) {\d \over {\d \vf _a (\vx)}} {\d \over {\d \vf _a
(\vy)}} + {{m c_A} \over 4\pi} \int \del \vf _a \bdel \vf _a + { c_A \over m\pi} \int
\del \vf _a (-\del \bdel)\bdel \vf _a  +... \eqno(8.5)
$$ Defining $\phi _a (\vk) = \sqrt {{c_A k \bar{k} }/ (2 \pi m)}~ \vf _a (\vk)$, we have
$$
\H \simeq \12 \int_x [- {\d ^2 \over {\d \phi _a ^2 (\vx)}} + \phi _a (\vx)  \bigl( m^2 -
\nabla ^2 \bigr)  \phi _a (\vx)] + ...
\eqno(8.6)
$$  We see that $\phi _a (\vx)$ behaves like a particle of mass $m$. (We also obtain the
relativistic energies $\sqrt{\vk ^2 + m^2}$ as mentioned at the end of the last
section.) We are currently investigating how $\O (\vf ^3)$-terms can correct these
results. 

The picture which emerges from our discussions is  as follows. We can think of $\phi_a
(\vx)$ as massive particles carrying non-Abelian charge.  When higher order terms are
included, clearly we will get an  interacting theory of these massive particles.
Although in the interest of finding an  eigenstate for $T$, we considered the special
choice of $f(\vx,\vy)$ in  section 6, with $\lambda ' \rightarrow 0, ~\lambda '\gg \e$,
we can, in Eq.(6.9), keep the separation $|\vx-\vy|$ finite and nonzero, which gives the
interaction $\V_{ca}(\vx,\vy)$ between the massive particles (and some other corrections
as well). We thus get a picture of the states being formed of massive constituents which
are interacting, the interaction binding them into states of zero charge. The $\phi_a$'s
are the ``constituents"  of the state. This is all in accord with the Schrodinger
equation we obtained at the end of the last section. It should be possible to develop
this constituent picture further, leading to a sequence of states as bound states of the
constituents with some interaction potential. This is under investigation.
\vskip .1in
\noindent{\bf 9. Discussion}
\vskip .1in

The mass was obtained by the action of the kinetic energy $T$ on $J^a$ and in this
context we consider the following potential counterargument to obtaining a mass at the
level of $T$ alone. The electric fields are the canonical momenta for $A^a (\vx)$ and
commute among themselves; so $T$ being $\int \vec{E}^2 /2$, we have a field theoretic
analogue of the free particle and would expect a continuous spectrum for $T$ alone. In
particular we could use an $\vec{E} $-diagonal representation with $E^a | f  \ra  = f^a
(\vx) | f \ra $, where $f$ is arbitrary, and hence $T$ can be made equal to any positive
number by choice of $f^a (\vx)$. Furthermore, Feynman has argued that one needs the
potential energy term to cut off possible ``escaping directions" in $ {\cal{A} /
\cal{G}} _*$, so that plane waves along such directions, which may have a continuous
spectrum, can be eliminated. We shall reexamine the ingredients which have gone into the
mass for $J^a (\vx)$ to see how these arguments are reconciled with our calculation.

In analyzing the ${\vec E}$-representation, it is useful to consider the following
parametrization of the electric fields. The complex component $E= \12 (E_1+iE_2)$
is an element of the Lie algebra of $SL(N,{\bf C})$
and therefore, except for a set of matrices of measure zero, it can be diagonalized by
a complex $SL(N,{\bf C})$-transformation $X$ [15]. Thus
$$
E= X~\Lambda ~X^{-1},~~~~~~~~~~~~~~~~~~~~{\bE}=X^{\dagger -1}~\Lambda^\dagger
~X^\dagger \eqno(9.1)
$$
where $\Lambda$ is a complex diagonal matrix.
In the ${\vec E}$-representation, $\Lambda$ and $X$ are $c$-numbers
and the gauge potentials $(A,\bA)$ become functional differential
operators as given by
$$\eqalign{
{\bar A}^a &= {i\over 2} D_{ab}(X) \left[ \sqrt{2} ~T^b_{ii} {\partial \over \partial \lambda_i}
-R^{bk} I_k\right]\cr
 A^a &= {i\over 2} D_{ba}(X^\dagger) \left[ \sqrt{2} ~T^a_{ii} {\partial \over \partial 
{\bar\lambda}_i}
-R^{*kb} {\bar I}_k\right]\cr
D_{ab}(X)&= 2~\Tr (T^a XT^b X^{-1})\cr
R^{ak} &= 2 \sum_{i\neq j} {T^a_{ij}T^k_{ji}\over {\lambda_i -\lambda_j}}\cr
[I^k, X]&= X~T^k,~~~~~~~~~~~~[{\bar I}^k, X^\dagger ] =T^k ~X^\dagger \cr}
\eqno(9.2)
$$
where $I,{\bar I}$ represent left and right translations on $X, X^\dagger$
respectively. In evaluating the action of $(A,{\bA})$ or the magnetic field
$B$ on a wave function in the ${\vec E}$-representation, the $R^{bk} I^k$ and
$R^{*kb}{\bar I}^k$ terms can bring in potential singularities when we 
have coincidence
of eigenvalues of $E$ due to the $(\lambda_i -\lambda_j )^{-1}$-factor.
(Notice that this factor and its contribution to $B$ via the commutator 
term are purely non-Abelian effects; they vanish for the Abelian theory.)
In particular, as ${\vec E} \rightarrow 0$, all eigenvalues tend to zero and the action of
the potential term on the wave function can become very large.
Although very explicit in the parametrization (9.1), this property
is simply a reflection of the uncertainty principle for $({\vec E},B)$ and is
not restricted to the specific parametrization.

Consider now a state with low values for the kinetic energy, say,
$\Psi ({\vec E})\sim \delta ( \int {\vec E}^2/2 -\epsilon )$ with $
\epsilon \rightarrow 0$. (There is also an additional phase factor
required by the implementation of the Gauss law in the 
${\vec E}$-representation.
We have not displayed this since it does not affect our arguments [16,17].)
As $\epsilon \rightarrow 0$, we need ${\vec E} \rightarrow 0$ since
$\int {\vec E}^2$ has a positive integrand. In this case, the contribution of
$B^2$ to the energy can become very large. Thus $T$ cannot be made arbitrarily small
keeping finiteness of the expectation value for $B^2$. 
Lowering the total energy requires some sort
of balance between the kinetic and potential energies and 
this could lead to a gap. In particular
for states with finite total energy, $B^2$ will have a finite expectation value. 
This argument
is still far from giving an understanding of our results in the
${\vec E}$-representation, but it does, we believe, carry the essential
physics of the problem. (The potential singularity for ${\vec E} \rightarrow 0$ can be
avoided for states for which the wave functions vanish near ${\vec E}=0$. The 
probability density for such states will be very small for small values of 
${\vec E}$ and hence there will be significant probability for  
finite nonzero values of
${\vec E}^2$. Therefore they can actually contribute a noninfinitesimal value to $T$.
Such states are not relevant to the potential counterargument which needs 
infinitesimal values for $T$.)

It may seem somewhat puzzling in this regard that we find a gap by 
considering $T$ alone, rather than $T+V$ as in the above argument.
Actually, we do have finite values for $\la B^2 \ra$. We are considering
states with finite norm where the inner product carries the factor 
$e ^{2 c_A \S (H)}$. Such states have finite expectation values for
$B^2 \sim : \bdel J_a (\vx) \bdel J_a (\vx) :$. In other words we are looking for
eigenstates of $T$ within the set of states of finite norm (and finite
$\la B^2 \ra$) and hence it is consistent with the previous argument
to get a mass gap even if the potential term is not included as part of the Hamiltonian.
In other words, it is not necessary to consider the spectrum of $ \int
\bigl( e^2 {\vec{E}^2 \over 2} + { B^2 \over {2 e^2}} \bigr) $ as a whole. 
It would be
possible to see a mass gap with $T$ alone provided the restriction to states of finite
$\la B^2 \ra $ arises via the inner product as in our case.
It may indeed be possible to obtain a continuous spectrum with no gap
for $T$ if we give up finiteness of $\la B^2 \ra$. Even if this may be a mathematical
possibility, it is clearly unphysical since we do eventually have to include
the potential term in the Hamiltonian and would need it to be finite.

\vskip .1in
\noindent{\bf Acknowledgements}
\vskip .1in This work was supported in part by the National Science Foundation grant
PHY-9322591 and by the Department of Energy grant DE-FG02-91ER40651-Task B.  C.K.'s work
was supported in part by the Korea Science and Engineering Foundation through the SRC
program. We thank B. Sakita for many useful discussions. C.K. also thanks the Physics
Department of the City College of the CUNY where part of the work was done. 

We thank R. Jackiw and the referee for raising questions which led to the discussion of
section 9.

\vskip .1in
\noindent{\bf References}
\vskip .1in
\item{1.} N. Seiberg, {\it Nucl.Phys.} {\bf B435} (1995); {\it Phys. Rev.} {\bf D49}
(1994) 6857; N. Seiberg and E. Witten, {\it Nucl.Phys.} {\bf B426} (1994) 19; {\it
ibid.} {\bf B430} (1994) 486; {\it ibid.} {\bf B431} (1994) 484; for a recent review,
see K. Intriligator and N. Seiberg, {\it Nucl.Phys. Proc.Suppl.} {\bf 45BC} (1996) 1.
\item{2.} D. Karabali and V.P. Nair, {\it Nucl.Phys.} {\bf B464} (1996) 135; {\it Phys.
Lett.} {\bf B379} (1996) 141; {\it Int. J. Mod. Phys.} {\bf A12} (1997) 1161. 
\item{3.}  A.M. Polyakov, {\it Nucl.Phys.} {\bf B120 } (1977) 429; G.'t Hooft, {\it
Nucl.Phys.} {\bf B138} (1978) 1; R. Jackiw and S. Templeton, {\it Phys.Rev.} {\bf D23}
(1981) 2291; J. Schonfeld, {\it Nucl.Phys.} {\bf B185} (1981) 157; S. Deser, R. Jackiw
and S. Templeton, {\it Phys.Rev.Lett.} {\bf 48} (1982) 975; {\it Ann.Phys.} {\bf 140}
(1982) 372; J. Goldstone and R. Jackiw, {\it Phys.Lett.} {\bf 74B} (1978) 81;  S.R. Das
and S. Wadia, {\it Phys.Rev.} {\bf D53} (1996) 5856; G. Grignani, G. Semenoff, P. Sodano
and O. Tirkkonen, UBC-GS-96-3, hep-th/9609228.
\item{4.} I.M. Singer, {\it Phys.Scripta} {\bf T24} (1981) 817;  {\it Commun. Math.
Phys.} {\bf 60} (1978) 7; M. Atiyah, N. Hitchin and  I.M. Singer, {\it Proc. Roy. Soc.
Lond.} {\bf A362} (1978) 425; P.K. Mitter and C.M. Viallet, {\it Phys. Lett.} {\bf B85}
(1979) 246; {\it Commun. Math. Phys.} {\bf 79} (1981) 457; M. Asorey and P.K. Mitter, 
{\it Commun. Math. Phys.} {\bf 80} (1981) 43; O. Babelon and C.M. Viallet,  {\it Commun.
Math. Phys.} {\bf 81} (1981) 515; {\it Phys. Lett.} {\bf B103} (1981) 45; P. Orland,
NBI-CUNY preprint, hep-th 9607134.
\item{5.} M.B. Halpern, {\it Phys. Rev.} {\bf D16} (1977) 1798; {\it ibid.}  {\bf D16}
(1977) 3515; {\it ibid.} {\bf D19} (1979) 517;  I. Bars and F. Green, {\it Nucl. Phys.}
{\bf B148} (1979) 445; D.Z. Freedman, {\it et al}, MIT preprint hep-th/9309045;  D.Z.
Freedman and R. Khuri, {\it Phys.Lett.} {\bf 192A} (1994) 153; M. Bauer and D.Z.
Freedman, {\it Nucl.Phys.} {\bf B450} (1995) 209; F.A. Lunev, {\it Phys.Lett.} {\bf
295B} (1992) 99; {\it Theor.Math.Phys.} {\bf 94} (1993) 66; {\it Mod.Phys.Lett} {\bf A9}
(1994) 2281; M. Asorey, {\it et al}, {\it Phys.Lett.} {\bf B349} (1995) 125;  O. Ganor
and J. Sonnenschein, {\it Int.J.Mod.Phys.} {\bf A11} (1996) 5701. 
\item{6.}  A.M. Polyakov and P.B. Wiegmann, {\it Phys.Lett.} {\bf B141}  (1984) 223; D.
Gonzales and A.N. Redlich, {\it Ann.Phys.(N.Y.)}  {\bf 169} (1986) 104; B.M. Zupnik,
{\it Phys.Lett.} {\bf B183} (1987) 175;  G.V. Dunne, R. Jackiw and C.A. Trugenberger,
{\it Ann.Phys.(N.Y.)} {\bf 149} (1989) 197; D. Karabali, Q-H Park, H.J. Schnitzer and Z.
Yang, {\it Phys.Lett.} {\bf B216} (1989) 307; D. Karabali and H.J. Schnitzer, {\it
Nucl.Phys.} {\bf B329} (1990) 649.
\item{7.}  E. Witten, {\it Commun.Math.Phys.} {\bf 92} (1984) 455; S.P. Novikov, {\it
Usp.Mat.Nauk} {\bf 37} (1982) 3.
\item {8.} K. Gawedzki and A. Kupiainen, {\it Phys.Lett.} {\bf B215} (1988) 119; {\it
Nucl.Phys.} {\bf B320} (1989) 649.
\item{9.} M. Bos and V.P. Nair, {\it Int.J.Mod.Phys.} {\bf A5} (1990) 959.
\item{10.} A. Migdal, {\it Zh.Eksp.Teor.Fiz.} {\bf 69} (1975) 810  ({\it Sov.Phys.JETP}
{\bf 42} (1975) 413); B. Rusakov, {\it Mod.Phys.Lett.} {\bf A5} (1990) 693; E. Witten,
{\it Commun.Math.Phys.} {\bf 141} (1991) 153; D. Fine, {\it Commun.Math.Phys.} {\bf 134}
(1990) 273; M. Blau and G. Thompson, {\it Int.J.Mod.Phys.} {\bf A7} (1992) 3781; D.
Gross, {\it Nucl.Phys.} {\bf B400} (1993) 161; D. Gross and W. Taylor IV, {\it Nucl.
Phys.} {\bf B400} (1993) 181; J. Minahan, {\it Phys.Rev.} {\bf D47} (1993) 3430.
\item{11.} V. Knizhnik and A.B. Zamolodchikov, {\it Nucl.Phys.} {\bf B247} (1984) 83.
\item{12.} N.N. Lebedev, {\it Special Functions and their Applications}, Dover
Publications, Inc. (1972) (English translation by R.A. Silverman).
\item{13.} V. Gribov, {\it Nucl.Phys.} {\bf B139} (1978) 1; I.M. Singer, {\it
Commun.Math.Phys.} {\bf 60} (1978) 7; T. Killingback and E.J. Rees, {\it
Class.Quant.Grav.} {\bf 4} (1987) 357.
\item{14.} J. Goldstone and R. Jackiw, {\it Phys. Rev.} {\bf D11} (1975) 1486; R.
Jackiw, {\it Rev. Mod. Phys.} {\bf 49} (1977) 681; J.L. Gervais and B. Sakita, {\it
Phys. Rev.} {\bf D11} (1975) 2943; J.L. Gervais, A. Jevicki and B. Sakita, {\it Phys.
Rev.} {\bf D12} (1975) 1038; N. Christ and T.D. Lee, {\it Phys. Rev.} {\bf D12} (1975)
1606.
\item{15.} J. Ginibre, J. Math. Phys. {\bf 6} (1965) 440.
\item{16.} M. Bauer, D.Z. Freedman and P.E. Haagensen, {\it Nucl. Phys.} {\bf B428}
(1994) 147; M. Bauer and D.Z. Freedman, ref. 5.
\item{17.} R. Jackiw, Talk at the AMS meeting, New York 1996, hep-th 9604040.

\end